\documentclass[12pt]{article}
\pdfoutput=1
\usepackage[pdftex]{graphicx}
\usepackage{amssymb}

\begin{document} 

\begin{titlepage}
	\rightline{}
	
	
	\vskip 2cm 
	\begin{center}
		\Large{{\bf From Black Hole to Qubits:\\Evidence of Fast Scrambling in BMN theory}}
	\end{center}
	
	\vskip 2cm 
	\begin{center}
		{Samuel Pramodh\footnote{\texttt{spramodh@g.hmc.edu}}\ \ \ and\ \ \ Vatche Sahakian\footnote{\texttt{sahakian@hmc.edu}}}\\
	\end{center}
	\vskip 12pt 
	\centerline{\sl Harvey Mudd College} 
	\centerline{\sl Physics Department, 241 Platt Blvd.}
	\centerline{\sl Claremont CA 91711 USA}
	
	\vskip 1cm 
	\begin{abstract}
		BMN Matrix theory admits vacua in the shape of large spherical membranes. Perturbing around such vacua, the setup provides for a controlled computational framework for testing information evolution in Matrix black holes. The theory realizes excitations in the supergravity multiplet as qubits. These qubits are coupled to matrix degrees of freedom that describe deformations of the spherical shape of the membrane. Arranging the ripples on the membrane into a heat bath, we use the qubit system as a probe and compute the associated Feynman-Vernon density matrix at one loop order. This allows us to trace the evolution of entanglement in the system and extract the characteristic scrambling timescale. We find that our numerical analysis is consistent with this time scaling logarithmically with the entropy of the qubit system, in tune with suggestions by Sekino and Susskind.
	\end{abstract}
\end{titlepage}

\newpage \setcounter{page}{1} 
\section{Introduction and Highlights}
\label{sub:intro}

Tracing the evolution of information falling into a black hole has recently become the focus of increased research activity. The premise naturally tests the limits of quantum information theory and string theory, and necessitates a peek into black hole horizon mechanics. It is now believed that, in one way or another, information falling past the black hole horizon gets deeply and quickly entangled with the degrees of freedom of the black hole~\cite{Maldacena:2001kr,Balasubramanian:2011ur,Balasubramanian:2011dm}. This process of information scrambling is not well understood. Complex physical systems scramble information over times scaling as a power law in the entropy. This phenomenon is generic, dictated by ergodic, random walk-like dissipation of information. Motivated by various paradoxes and thought experiments, the authors of~\cite{Sekino:2008he,Susskind:2011ap} have proposed that a black hole is the fastest scrambling machine, diffusing information on timescales given by
\begin{equation}\label{eq:main}
	\tau \sim \frac{\ln S}{T}\ ,
\end{equation}
where $S$ is entropy and $T$ is temperature.
Indeed, any faster scrambling would lead to a violation of the no-cloning principle of quantum mechanics. To date, there is no known physical system proven to achieve such a fast rate of scrambling. Contrived non-physical models have been developed in the literature~\cite{Lashkari:2011yi,Hayden:2007cs}, but none that are related to black holes or gravitational dynamics, let alone being realized within string theory (other interesting efforts in this direction can be found in~\cite{Barbon:2011nj}-\cite{Iizuka:2013kha}). If the conjecture of~\cite{Sekino:2008he} is correct, string theory -- or any candidate theory of quantum gravity --  should be able to demonstrate this fast scrambling phenomenon. In~\cite{Susskind:2011ap,Asplund:2011qj}, it was suggested that Matrix theory, as a framework for M theory, may provide for a fast scrambling mechanism. 

In a previous work~\cite{Brady:2013opa}, a controlled setting for addressing the fast scrambling conjecture was developed within Berenstein-Maldacena-Nastase (BMN) Matrix theory~\cite{Berenstein:2002jq}. The latter is related to the Banks-Fischler-Shenker-Susskind (BFSS) Matrix model~\cite{Banks:1996vh} and is also conjectured to describe dynamics of M-theory -- in particular light-cone M-theory in a plane-wave background. The BMN setup comes with a infra-red cutoff (related to the background plane wave) which allows for vacua in the form of Bogomol'nyiâ-Prasad-Sommerfield (BPS) spherical membranes (M2 branes of light-cone M-theory), or fuzzy spheres (D2 branes of IIA theory)~\cite{Dasgupta:2002hx,Ling:2006up}. In the proper regime of the parameters, gravitational geometries have been identified that are dual to these Matrix vacua~\cite{Lin:2004nb}. On the Matrix theory side, one can perturb around these vacua and effectively study the dynamics of these objects. 

\begin{figure}
	\begin{center}
		\includegraphics[width=5.5in]{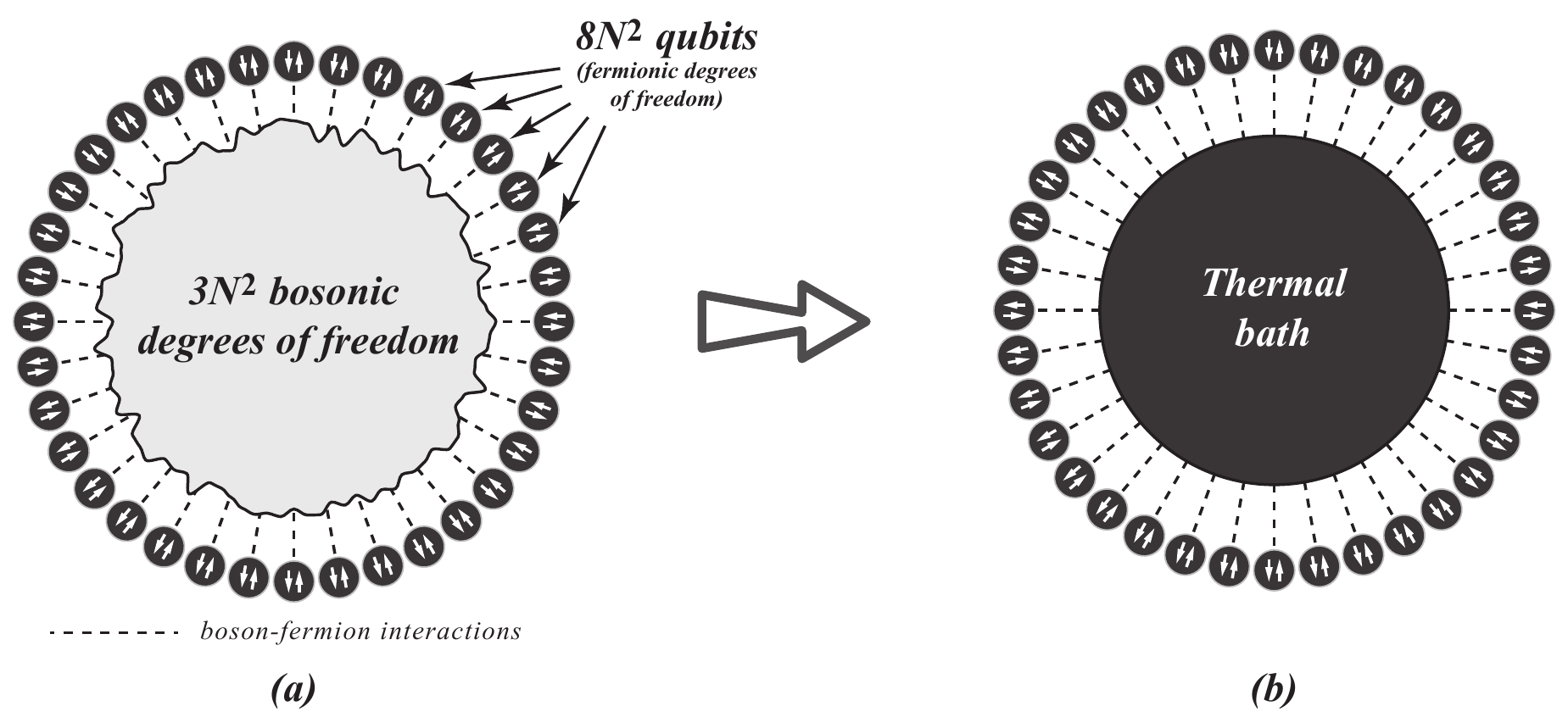} 
	\end{center}
	\caption{(a) A cartoon of the sector of the BMN Matrix theory dynamics of interest. $8\,N^2$ qubits correspond to the spherical harmonic modes of the eleven dimensional supergravity multiplet on a sphere. The system has both an infrared (IR) cutoff (the size of the sphere) and an ultraviolet (UV) cutoff (related to $N$, the regularization of the fuzzy sphere). The remaning $3\,N^2$ bosonic degrees of freedom describe the deformations of the shape of the M2 sphere in a three dimensional subset of the full space. (b)  Our setup makes a heat bath out of the membrane ripples, in the background of which the qubits evolve. Integrating out the bosonic heat bath generates a dense network of interactions between the qubits. We look for entanglement evolution in this qubit system to determine whether BMN Matrix theory has the right ingredients to be a fast scrambler.}\label{fig:cartoon}
\end{figure}
These perturbations of the fuzzy spheres come in two types: geometric deformations of the spherical membrane, appearing as bosonic degrees of freedom in the Matrix theory; and harmonics of the eleven dimensional supergravity multiplet -- arising from fermionic degrees of freedom in the Matrix theory. In~\cite{Brady:2013opa}, it was shown that, for the vacuum of BMN theory corresponding to a single spherical M2 brane with $N$ units of light-cone momentum, the perturbation dynamics can be mapped onto the dynamics of $8\,N^2$ qubits coupled to $3\,N^2$ bosonic variables that describe the shape of the M2 brane (see Figure~\ref{fig:cartoon}). It was then demonstrated that the coupling between the qubits and the bosonic variables is necessary if the BMN theory is to achieve fast scrambling in the qubit sector.

In this work, we focus on an explicit computation in BMN Matrix theory with aim to ascertain whether the theory is indeed a fast scrambler\footnote{Other works exploring the dynamics of BMN Matrix theory can be found in~\cite{Berenstein:2010bi,Asplund:2012tg,Riggins:2012qt}}. We start with the system of~\cite{Brady:2013opa} and arrange the bosonic degrees of freedom in a heat bath at a fixed temperature $T$. This represents for us a background Matrix black hole model~\cite{Horowitz:1997fr,Banks:1997hz,Banks:1997tn} through which we can trace the evolution of the qubits. We then compute the density matrix of the qubit system and measure entanglement as a function of time at one loop order. Since the qubits represent the graviton, 3-form, and gravitino modes of eleven dimensional supergravity, we are then tracking the evolution of information stored in the supergravity multiplet near the surface of a spherical rippling M2 brane. We are able to obtain a closed form expression for the entanglement entropy. We then extract the timescale of information scrambling from this result. Our results are consistent with a scrambling time that scales as the logarithm of the entropy of the qubits.

In section 2, we develop the perturbation theory about the BPS membrane vacua of BMN theory. In Section 3, we present the core technology, the Feynman-Vernon technique of integrating out a heat bath; and we apply it to our system leading up to identifying the fast scrambling timescale. In Section 4, we present our conclusions and a discussion of future directions. The Appendices collect some of the more detailed results: Appendix A shows samples of combinatorial terms arising from the Feynman diagrams used in the computation; Appendix B shows a justification of the chosen coupling constant for a perturbative expansion; and Appendix C shows the main result -- part of the computed entanglement entropy as a function time -- used in identifying fast scrambling.

\section{Setup}

The BMN system is a $0+1$ dimensional supersymmetric Matrix theory describing the dynamics of $N$ D0 branes. The Lagrangian is given by~\cite{Berenstein:2002jq,Dasgupta:2002hx}
\begin{eqnarray}
	L&=\frac{1}{2}R\,\mbox{Tr} \left[
	\frac{1}{R^2}\left(D_t X_i\right)^2+\frac{1}{R^2}\left(D_t Y_a\right)^2 + \frac{1}{2} \left[
	X_i, X_j\right]^2+\left[X_i, Y_a\right]^2+\frac{1}{2}\left[Y_a,Y_b\right]^2 \right. \nonumber \\
	&- \left. \left(\frac{\mu}{3 R}\right)^2 X_i^2-\left(\frac{\mu}{6 R}\right)^2 Y_a^2-\frac{i}{3} \frac{\mu}{R}\, \epsilon_{ijk} \left[X_i,X_j\right] X_k \right. \nonumber \\
	&+ \left. \frac{1}{R}\Psi D_t \Psi + \Psi \gamma_i \left[X_i, \Psi\right]+\Psi \gamma_a \left[Y_a,\Psi\right]-\frac{i}{4} \frac{\mu}{R} \Psi \gamma_{123} \Psi
	\right]\ .
\end{eqnarray}
The $X_i$s and $Y_a$s are $N\times N$ Hermitian matrices with $i=1,2,3$ and $a=4,\ldots, 9$. The $\Psi$ is also a matrix in the adjoint of $U(N)$ with entries that are Majorana-Weyl spinors in ten dimensions. Two variables parameterize the system, $R$ and $\mu$. The limit where $\mu\rightarrow 0$ leads to the BFSS Matrix model, that is D0 branes in a flat background or M theory in the light cone frame with light cone momentum $p^+=N/R$. For $\mu\neq 0$, the theory describes D0 branes in a plane wave background, or light-cone M-theory in a plane wave background with again $p^+=N/R$.
We fix the static gauge $A_t=0$, with the covariant derivative $D_t$ becoming simply the time derivative -- at the cost of the constraint
\begin{equation}
	i [X^i, \Pi^i]+i [Y^a, \Pi^a] + i \Psi\Psi = 0\ ,
\end{equation}
where the $\Pi$s are the canonical momenta of the $X$s.
In our conventions, the length units of the various variables are given by
\begin{equation}
	[X],[Y],[\Psi]\sim 1\ \ ,\ \ [t]\sim L\ \ ,\ \ [R]\sim L^{-1}\ \ ,\ \ [\mu]\sim L^{-1}\ .
\end{equation}

The BPS vacua are described by the configuration
\begin{equation}
	X_i=\frac{\mu}{6\, R} \tau_i\ \ \ ,\ \ \ Y_a = 0\ \ \ ,\ \ \ \Psi=0
\end{equation}
The $\tau_i$s are the Pauli matrices, with $[\tau_i,\tau_j]=2\,i\, \varepsilon_{ijk}\tau_k$. Considering $N$ dimensional representation of $SU(2)$, these vacua are fuzzy (sometimes called non-commutative) spheres of radii\footnote{In the language of~\cite{Ling:2006up}, this corresponds to $N_2=1$ and $N=N_5\gg 1$ for a single spherical membrane. The theory can also describe spherical M5 branes in other regimes of the parameters~\cite{Ling:2006up}.}
\begin{equation}\label{eq:radius}
	r=\sqrt{\frac{\mbox{Tr} X_i^2}{N}} = \frac{\mu}{6\, R} {\frac{\sqrt{N^2-1}}{\sqrt{3}}}\sim \frac{\mu N}{R}\ \ \ \mbox{for large $N\gg 1$}\ .
\end{equation}
This setup is our starting point. In a proper regime of the paramaters and at finite temperature, it can serve as a model for a black hole within a fully quantized theory of gravity. We proceed by perturbing the vacuum as follows
\begin{equation}
	X_i=\frac{\mu}{6\, R} \tau_i+x_i \sqrt{\frac{R}{\mu}}\ \ \ ,\ \ \ Y_a = 0\ \ \ ,\ \ \ \Psi=\psi\ .
\end{equation}
We freeze out the perturbations in the six transverse directions, $Y_a=0$, essentially focusing on a compactification to three target space dimensions. We also rescale time
\begin{equation}
	t\rightarrow \frac{t}{\mu}\ .
\end{equation}
Henceforth, all variables are then dimensionless, and time or energy are measured in units of $\mu$. 

We then expand all matrix structures in matrix spherical harmonics $Y^j_m$ with $j=1,\ldots,N-1$ and $m=-j,\ldots,j$~\cite{Dasgupta:2002hx,reza}: note that we have~\footnote{The $j=0$ case corresponds to the center of mass degree of freedom -- the $U(1)$ in $U(N)$.}
\begin{equation}
\sum_{j=0}^{N-1} (2j+1) = \mbox{dim}[U(N)]
\end{equation}
as expected. And the $Y^j_m$'s are $N\times N$ matrices satisfying the algebra~\cite{hoppe}
\begin{equation}
	\left[
		Y^{j}_{m},Y^{j'}_{m'}
	\right] = \frac{2}{N} \sqrt{(2j+1)(2j'+1)(2j''+1)} f^{jm\,j'm'}_{j''m''} (-1)^{m''} Y^{j''}_{-m''}
\end{equation}
where
\begin{equation}\label{eq:fs}
	f^{jm\,j'm'}_{j''m''} = (-1)^N N^{3/2} \times \left(
	\begin{array}{ccc}
		j & j' & j'' \\
		m & m' & m''
	\end{array}
	\right)\times
	\left\{
	\begin{array}{ccc}
		j & j' & j'' \\
		\frac{N-1}{2} & \frac{N-1}{2} & \frac{N-1}{2}
	\end{array}
	\right\}
\end{equation}
written in terms of 3$j$ and 6$j$ symbols. We also have the normalization condition 
\begin{equation}
	\mbox{Tr} \left(
	Y^j_m Y^{j'}_{m'}
	\right) = (-1)^m N \delta_{jj'} \delta_{-mm'}
\end{equation}
For the bosonic variables, the decomposition that diagonalizes the quadratic part of the Hamiltonian looks like~\cite{Dasgupta:2002hx}
\begin{eqnarray}
	&& x_1= \sum_{j,m}
	\frac{Y^j_m}{2 \sqrt{N (2 j+1)}} \times \nonumber \\ &&\left[ \left({\frac{\sqrt{(j-m-1) (j-m)}}{\sqrt{j}} \alpha _{j-1\,m+1}-\frac{\sqrt{(j+m-1) (j+m)}}{\sqrt{j}} \alpha _{j-1\,m-1}}\right) \right. \nonumber \\
	&&+ \left. \left({\frac{\sqrt{(j-m+1) (j-m+2)}}{\sqrt{j+1}} \beta_{j+1\,m-1}-\frac{\sqrt{(j+m+1) (j+m+2)}}{\sqrt{j+1}} \beta _{j+1\,m+1}}\right)\right]
\end{eqnarray}
\begin{eqnarray}
	&&x_2= \sum_{j,m}
	\frac{Y^j_m}{2 \sqrt{N (2 j+1) }} \times \nonumber \\ && \left[ i \left({\frac{\sqrt{(j+m-1) (j+m)}}{\sqrt{j}} \alpha _{j-1\,m-1}+\frac{\sqrt{(j-m-1) (j-m)}}{\sqrt{j}} \alpha _{j-1\,m+1}}\right) \right. \nonumber \\
	&&+\left. i \left({-\frac{\sqrt{(j-m+1) (j-m+2)}}{\sqrt{j+1}} \beta_{j+1\,m-1}-\frac{\sqrt{(j+m+1) (j+m+2)}}{\sqrt{j+1}} \beta _{j+1\,m+1}}\right)\right]
\end{eqnarray}
\begin{equation}
	x_3= \sum_{j,m} \frac{Y^j_m}{\sqrt{{N (2j+1)}}}  \times \left[
	{\sqrt{\frac{j^2-m^2}{j}} \alpha _{j-1\,m}+\sqrt{\frac{(j-m+1) (j+m+1)}{(j+1)}} \beta _{j+1\,m}}
	\right]\ .
\end{equation}
The bosonic degrees of freedom are hence written in terms of physical complex variables $\alpha_{jm}(t)$ and $\beta_{jm}(t)$. This decomposition diagonalizes the quadratic part of the Lagrangian. The indices $j$ and $m$ are summed over such that the range of the indices on the $\alpha_{jm}$s is $j=1,\ldots,N-2$ and $m=-j,\ldots,j$, while the range on the indices of the $\beta_{jm}$s is $j=1,\ldots,N$ and $m=-j,\ldots,j$.

The Majorana-Weyl spinors of $SO(9,1)$ can be conveniently written as spinors of $SO(3)\times SO(6)\sim SU(2)\times SU(4)$. We then write each of the two components of the fundamental of the $SU(2)$ in terms of $SU(4) $ spinors $\eta_{jm}(t)$ (fundamental) and $\chi_{jm}(t)$ (anti-fundamental)~\cite{Dasgupta:2002hx}
\begin{equation}
	\Psi=\sum_{j,m} \frac{Y^j_m}{\sqrt{(2 j+1) {N}}} \left(
	\begin{array}{c}
		{\sqrt{j+m+1} \eta _{j+\frac{1}{2}\,m+\frac{1}{2}}-\sqrt{j-m} \bar{\chi }_{j-\frac{1}{2}\,m+\frac{1}{2}}} \\
		{\sqrt{j+m} \bar{\chi }_{j-\frac{1}{2}\,m-\frac{1}{2}}+\sqrt{j-m+1} \eta _{j+\frac{1}{2}\,m-\frac{1}{2}}}
	\end{array}
	\right)\ .
\end{equation}
This is once again arranged to diagonalize the quadratic part of the Lagrangian. The summed indices $j$ and $m$ are such that the range on the $\eta_{jm}$ indices is $j=1/2,\ldots,N-(1/2)$ and $m=-j,\ldots,j$ while the range on the $\chi_{jm}$ indices is $j=1/2,\ldots,N-(3/2)$ and $m=-j,\ldots,j$.

Note that due to the conjugation property
\begin{equation}
	(Y^{j}_{m})^\dagger = (-1)^m Y^{j}_{-m}\ .
\end{equation}
the $\alpha_{jm}$s and $\beta_{jm}$s satisfy the conditions 
\begin{equation}
\alpha_{jm}^* = (-1)^m \alpha_{j\,-m}\ \ \ ,\ \ \ \beta_{jm}^* = (-1)^m \beta_{j\,-m}
\end{equation}
We can then easily check 
\begin{equation}
\underbrace{N\,(N-2)}_{\mbox{$\alpha$'s}}+\underbrace{N\, (N+2)}_{\mbox{$\beta$'s}}+\underbrace{N^2 \vphantom{|}}_{\mbox{$U(N)$ \rlap{gauge dir.}}}=3\,N^2\ .
\end{equation}
as required. Hence, the physical degrees of freedom are mapped onto all of the $\alpha_{jm}$s and $\beta_{jm}$s. Similarly we find $8\,N^2$ fermionic degrees of freedom amongst the $\eta_{jm}$s ($4\,N\,(N+1)$ in total) and $\chi_{jm}$s ($4\,N\,(N-1)$ in total) .
Substituting these decompositions into the Lagrangian, we obtain
\begin{equation}
	L=L_{BB}+L_{FF}+g\,L_{BBB}+g\,L_{BFF}+g^2 L_{BBBB}
\end{equation}
where $g$ is the effective coupling constant for large $N$
\begin{equation}\label{eq:g}
	g \equiv \left(\frac{R}{\mu N}\right)^{3/2}
\end{equation}
and the notation is such that $L_{BBB}$ is cubic in the bosonic variables, $L_{BFF}$ is linear in the bosonic variables and quadratic in the fermionic ones, etc. $L_{BBB}$ and $L_{BFF}$ are linear in the structure constants $f^{jm\,j'm'}_{j''m''}$, while $L_{BBBB}$ is quadratic. In Appendix B, we demonstrate that these $f^{jm\,j'm'}_{j''m''}$s are $N$ independent for large $N$. That is, all explicit $N$ dependence in the Lagrangian have been factored into our definition of the effective coupling constant $g$.

For small $g$ and large but finite $N$, we have a well-defined perturbative expansion in $g$. The radius of the membrane scales as $g^{-2/3}$ as seen from~(\ref{eq:radius}). This assures that the membrane is large in Planck units. However, for $g\,N\sim 1$, we can instead view the setup as a spherical D2 brane with a non-commutative worldvolume theory~\cite{Ling:2006up}. From this perspective, the theory has an IR cutoff set by $\mu$. The scale of non-commutativity is given by
\begin{equation}
	\theta = \frac{1}{\mu^2 N}\ .
\end{equation} 
And $N$ tunes the ratio of the IR and UV cutoffs. The dimensionless coupling on the worldvolume is given by
\begin{equation}
	g_{eff}^2(\mu)=\frac{g_{YM}^2}{\mu^2} = g^2 N^2\ ,
\end{equation}
at the IR scale. The temperature in the system, $1/\beta$, should be between the IR and UV cutoffs
\begin{equation}
	\frac{1}{N} \lesssim \beta \lesssim 1\ ,
\end{equation}
where temperature is measured in units of $\mu$. And the effective dimensionless coupling at the temperature scale of interest is then $g_{eff}^2(1/\beta) = g^2 N^2 \beta^2$. Thus, for
\begin{equation}\label{eq:regime}
	g \ll 1\ \ \ ,\ \ \ N \gg 1\ \ \ ,\ \ \ \frac{1}{g\,N}<\beta \lesssim 1\ \ \ ,\ \ \ g\,N> 1\ ,
\end{equation}
we can capture the {\em strong} coupling regime $g_{eff}^2(1/\beta)\gtrsim 1$ of this non-commutative D2 brane worldvolume dynamics through a perturbative expansion in $g$ in BMN Matrix theory\footnote{The perturbation regime is that of the commutative $2+1$ dimensional theory with worldvolume flux and the strongly coupled regime is that of the non-commutative version~\cite{Seiberg:1999vs}.}. To probe this interesting regime, we can choose for example $g\sim 1/10$, $\beta \lesssim 1$, and $10<N<100$. 

The reason we are interested in the regime delineated by~(\ref{eq:regime}) is that strongly coupled non-commutative dynamics of a spherical D2 brane is rather reminiscent of the Matrix black hole proposals of~\cite{Horowitz:1997fr,Banks:1997hz,Banks:1997tn}. Hence, through perturbation in $g$ we may expect to capture the essence of black hole horizon physics -- keeping in non-local dynamics in the picture yet scaling out superfluous complications from gravity, as is typical in non-commutative field theory settings~\cite{Seiberg:1999vs, Maldacena:1999mh}.

\subsection{Pure bosonic terms}

The bosonic sector of the dynamics involves properly diagonalized quadratic terms~\cite{Dasgupta:2002hx}
\begin{eqnarray}
	L_{BB}&=&\sum_{j=1}^{N-2} \sum_{m=1}^j\left|\dot{\alpha}_{jm}\right|^2-\left(\frac{j+1}{3}\right)^2\left|{\alpha_{jm}}\right|^2
	+\sum_{j=1}^{N} \sum_{m=1}^j\left|\dot{\beta}_{jm}\right|^2-\left(\frac{j}{3}\right)^2\left|{\beta_{jm}}\right|^2 \nonumber \\
	&+&\sum_{j=1}^{N-2} \frac{1}{2}\dot{\alpha}_{j0}^2-\frac{1}{2}\left(\frac{j+1}{3}\right)^2{\alpha_{j0}}^2+\sum_{j=1}^{N} \frac{1}{2}\dot{\beta}_{j0}^2-\frac{1}{2}\left(\frac{j}{3}\right)^2{\beta_{j0}}^2\ . \label{eq:LBB}
\end{eqnarray}
identifying two sets of masses in units of $\mu$. It will be useful to write these terms in a more compact notation where we package the $N\,(N-2)$ components of the $\alpha_{jm}$s as
\begin{equation}
	\mathcal{A}_j=\left(\underbrace{\Re \alpha_{j1},\cdots,\Re \alpha_{jj}}_{K=1},\underbrace{\Im \alpha_{j1},\cdots,\Im \alpha_{jj}}_{K=2},\underbrace{\alpha_{j0}}_{K=3}\right) \rightarrow \mathcal{A}^K_j
\end{equation}
with $j=1\,\ldots,N-2$ and $K=1,2,3$; and we package the $N(N+2)$ components of the $\beta_{jm}$s as
\begin{equation}
	\mathcal{B}_j=\left(\underbrace{\Re\beta_{j1},\cdots,\Re\beta_{jj}}_{K=1},\underbrace{\Im\beta_{j1},\cdots,\Im\beta_{jj}}_{K=2},\underbrace{\beta_{j0}}_{K=3}\right) \rightarrow \mathcal{B}^K_j
\end{equation}
where $j=1\,\ldots,N$ and $K=1,2,3$. Equation~(\ref{eq:LBB}) then takes the form
\begin{equation}
	L_{BB}=\sum_{j=1}^{N-2}\frac{1}{2}{\dot{{\mathcal{A}}}}^K_j\cdot {\dot{{\mathcal{A}}}}^K_j-\frac{1}{2} \left(\frac{j+1}{3}\right)^2 \mathcal{A}^K_j\cdot \mathcal{A}^K_j+\sum_{j=1}^{N}\frac{1}{2}{\dot{{\mathcal{B}}}}^K_j\cdot {\dot{{\mathcal{B}}}}^K_j-\frac{1}{2} \left(\frac{j}{3}\right)^2 \mathcal{B}^K_j\cdot \mathcal{B}^K_j
\end{equation}
where the dot represents sum over the $m$-tagged components, and with an implicit sum over $K=1,2,3$. 

The remaining terms of the Lagrangian, $L_{BBB}$ and $L_{BBBB}$, involve cubic and quartic interactions between these modes -- scaled respectively by $g$ and $g^2$. In the limit of small $g$ we will be focusing on, we will see that we do not need to consider these interactions for the purpose of demonstrating fast scrambling. Hence we do not show the explicit and rather complicated forms of $L_{BBB}$ and $L_{BBBB}$. 

\subsection{Fermionic terms}

The fermionic sector involves diagonalized quadratic terms~\cite{Dasgupta:2002hx}
\begin{eqnarray}
	L_{FF}&=&\sum_{j=1/2}^{N-1/2} \sum_{m=-j}^{j} i\, \overline{\eta}_{jm}\cdot \dot{\eta}_{jm}-\frac{1}{3} \left(j+\frac{1}{4}\right) \overline{\eta}_{jm}\cdot {\eta}_{jm} \nonumber \\
	&+&	\sum_{j=1/2}^{N-3/2} \sum_{m=-j}^{j} i\, \overline{\chi}_{jm}\cdot \dot{\chi}_{jm}-\frac{1}{3} \left(j+\frac{3}{4}\right) \overline{\chi}_{jm}\cdot {\chi}_{jm}\ ,\label{eq:LFF}
\end{eqnarray}
where the dot implies sum over the $SU(4)$ indices of the spinors. 
Quantizing this diagonalized form leads to the Hamiltonian
\begin{equation}\label{eq:HFF}
H_{FF}=\sum_{j=1/2}^{N-1/2} \sum_{m=-j}^{j} \frac{1}{3} \left(j+\frac{1}{4}\right) \overline{\eta}_{jm}\cdot {\eta}_{jm}+\sum_{j=1/2}^{N-3/2} \sum_{m=-j}^{j} \frac{1}{3} \left(j+\frac{3}{4}\right) \overline{\chi}_{jm}\cdot {\chi}_{jm} \ .
\end{equation}
with the commutation relations
\begin{equation}
 \{\overline{\eta}_{j'm'},\eta_{jm}\} = \delta_{jj'} \delta_{mm'}\ \ \ , \ \ \  \{\overline{\chi}_{j'm'}, \chi_{jm} \} = \delta_{jj'} \delta_{mm'}\ .
\end{equation}
Hence, we have a system of $8\,N^2$ disjoint qubits (see~\cite{Brady:2013opa} for details). However, there is also coupling to the bosonic sector through the terms
\begin{equation}
	L_{BFF}=g\,\sum_{j,K}\omega^{K}_{j}\cdot \mathcal{A}^K_j+g\,\sum_{j,K}\omega^{K+3}_{j}\cdot \mathcal{B}^K_j
\end{equation}
where we have used the compact notation introduced in the previous section; and the $\omega$s are defined by
\begin{eqnarray}
	\left(\omega^{K}_{j''}(t)\right)_{m''} &=& \sum_{j=1/2}^{N-1/2} \sum_{m=-j}^{j} \sum_{j'=1/2}^{N-1/2} \sum_{m'=-j}^{j'} \overline{\eta}_{jm}(t)\cdot \eta_{j'm'}(t) \left(W_{jmj'm';j''}^{(1)\,K}\right)_{m''} \nonumber \\
	&+&\sum_{j=1/2}^{N-3/2} \sum_{m=-j}^{j} \sum_{j'=1/2}^{N-3/2} \sum_{m'=-j}^{j'} \overline{\chi}_{jm}(t)\cdot \chi_{j'm'}(t) \left(W_{jmj'm';j''}^{(2)\,K}\right)_{m''} \nonumber \\
	&+&\sum_{j=1/2}^{N-3/2} \sum_{m=-j}^{j} \sum_{j'=1/2}^{N-1/2} \sum_{m'=-j}^{j'} {\chi}_{jm}(t)\cdot \eta_{j'm'}(t) \left(W_{jmj'm';j''}^{(3)\,K}\right)_{m''} \nonumber \\
	&+&\sum_{j=1/2}^{N-3/2} \sum_{m=-j}^{j} \sum_{j'=1/2}^{N-1/2} \sum_{m'=-j}^{j'} \overline{\chi}_{jm}(t)\cdot \overline{\eta}_{j'm'}(t) \left(W_{jmj'm';j''}^{(4)\,K}\right)_{m''}\ .
\end{eqnarray}
The coefficients shown as $W_{jmj'm';j''}^{(i)K}$, with $i=1,2,3,4$, are rather complicated numerical expressions. Samples are shown in Appendix A. They satisfy the identities
\begin{eqnarray}
	\left(W_{jmj'm';j''}^{(1)K}\right)_{m''} &=& \left(W_{j'm'jm;j''}^{(1)K}\right)_{m''}^*\ \ \ ,\ \ \ 
	\left(W_{jmj'm';j''}^{(2)K}\right)_{m''} = \left(W_{j'm'jm;j''}^{(2)K}\right)_{m''}^*\nonumber \\
	\left(W_{jmj'm';j''}^{(3)K}\right)_{m''} &=& -\left(W_{jmj'm';j''}^{(4)K}\right)_{m''}^*
\end{eqnarray}
which assure that the Lagrangian is real. They consist of linear combinations of the structure constants $f^{jm\,j'm'}_{j''m''}$ of equation~(\ref{eq:fs}), and hence involve the combinatorics of combining angular momentum modes through $3j$ and $6j$ symbols. Appendix B shows numerical profiles of these structure constants, demonstrating that the $f^{jm\,j'm'}_{j''m''}$s scale as $N^0$ for large $N$. All explicit $N$ dependence of the action is hence factored into our definition of the coupling constant $g$ through equation~(\ref{eq:g}).

\subsection{Perturbations}

For weak coupling $g\ll 1$, the fermionic sector consists of $8\,N^2$ qubit coupled to the bosonic sector through the $L_{BFF}$ term. The maximum entropy of the qubits scales as $N^2$. We want to see whether the coupling between the fermionic and bosonic sectors can lead to the fast scrambling expected of a quantum theory of gravity as proposed in~\cite{Sekino:2008he}, with scrambling timescale $\tau\sim \ln N/T$.

\begin{figure}
	\begin{center}
		\includegraphics[width=6.5in]{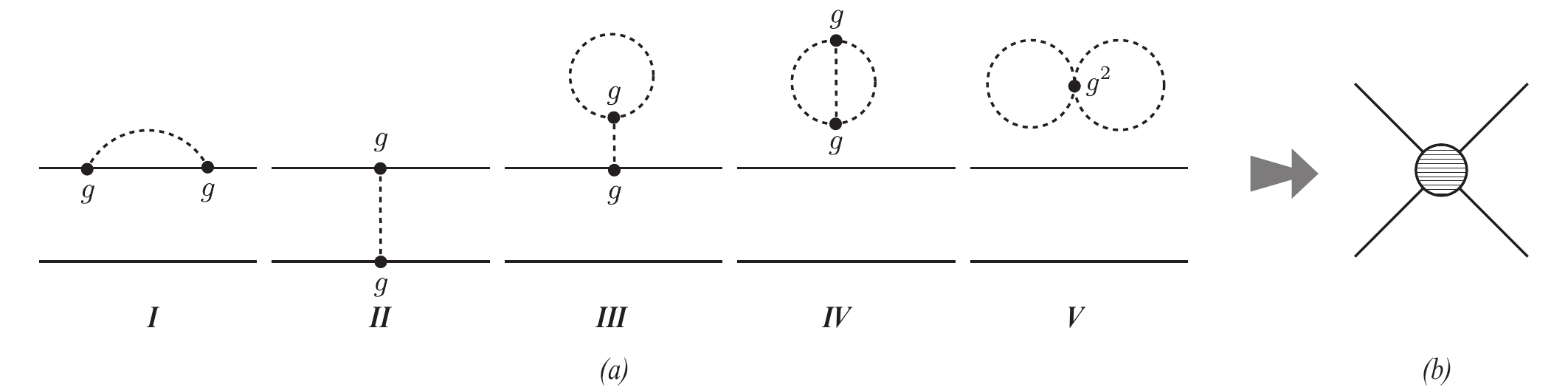} 
	\end{center}
	\caption{(a) Configuration space Feynman diagrams that contribute to our computation, to leading order in $g^2$. The solid lines represent qubit states while the dotted lines represent the bosonic degrees of freedom (arranged in a heat bath). Diagrams I, II, and III involve the $L_{BFF}$ terms from the Lagrangian. Diagrams III and IV involve the $L_{BBB}$ terms. While diagram V involves the $L_{BBBB}$ terms. Each line has $\sim N^2$ polarizations labeled by the angular momenta indices $j$ and $m$. Thus, a vertex with three prongs potentially corresponds to $\sim N^6$ Feynman rules while a quartic vertex would correspond to $\sim N^8$ rules. Each three prong vertex is proportional to a structure constant $f$ from~(\ref{eq:fs}) -- essentially a Clebsch–Gordan coefficient that couples the various angular momentum modes -- while the four prong vertex involves two $f$s. (b) Integrating out the bosonic heat bath leaves a Fermi four-qubit interaction that creates a large number of interaction links between the $8\,N^2$ qubits of the system. The scrambling mechanism arises entirely from diagrams I, II, and III, all involving one or more $L_{BFF}$ couplings. This high density of effective qubit interactions originates from the fact that many $f$s of~(\ref{eq:fs})  are non-zero -- {\em i.e.} an angular momentum mode couples to almost all other angular momentum modes. Interestingly, the strongest couplings arise between the largest and smallest angular momentum modes.}\label{fig:diagrams}
\end{figure}

Our approach is to compute the density matrix for the qubits as a function of time  -- using the bosonic sector as a heat bath at a fixed temperature. We then construct the reduced density matrix by integrating over half of the qubits and obtain the entanglement entropy as a function of time. Analyzing this, we extract the characteristic scrambling timescale as a function of $N$.  To manage the computation, we consider the limit of small coupling constant $g\ll 1$ and large $N\gg 1$.
In this small $g$ regime, our leading order computation amounts to considering the schematic Feynman diagrams shown in Figure~\ref{fig:diagrams}. We would then be integrating out the bosonic dynamics to order $g^2$ leading to a four qubit Fermi interaction. These quartic interactions are bound to generate a dense network of connections between the qubits, and the question we are to address is whether this high density of connections is enough to lead to {\em fast} scrambling of qubit information. In this regard, note that entanglement can only be generated by a boson-fermion-fermion vertex since this is the only dynamical mechanism for flipping and entangling qubits. In particular, there is no direct scrambling at a work in the last two diagrams of Figure~\ref{fig:diagrams}(a).

\section{Density matrix}

\subsection{Setup}

Our first goal is to compute the density matrix of the qubit system in the background of a bosonic thermal bath. We do this by using the Feynman-Vernon path integration technique~\cite{Feynman:1963fq}. In this language, the $\eta_{jm}$s and $\chi_{jm}$s become Grassmanian variables. To simplify the notation, we represent them as $F$
\begin{equation}
	F\rightarrow \eta_{jm}^{I}\mbox{   or   } \chi_{jm}^{I}\ ,
\end{equation}
with $I$ being the $SU(4)$ index.
The $F$s then satisfy
\begin{equation}
\{F, \overline{F}\} = 0\ .
\end{equation}
The bosonic variables are similarly written as 
\begin{equation}
B\rightarrow \alpha_{jm}\mbox{   or   } \beta_{jm}\ .
\end{equation}
In subsequent path integral expressions, $F$ and $B$ represent {\em all} of the fermionic and bosonic degrees of freedom.

\subsection{Integrating out the thermal bath}

The novelty that arises in the Feynman-Vernon technique -- as opposed to the traditional path integration approach to scattering amplitude computations -- is that the in-out qubits (row-column entry labels) of the density matrix specify the boundary conditions in the path integration. In addition, we need to adapt the original Feynman-Vernon technique to a system involving fermions. As depicted in Figure 1(a), the computation is expected to be at the one-loop level to order $g^2$. 
Our starting point is the Feynman-Vernon density matrix, properly adapted to Grassmanian variables and using a measure consistent with the Grassmanian coherent state formalism~\cite{Grosche:1998yu}
\begin{equation}
	\rho(F_f,F'_f,t) = \int d\overline{F}_i dF_i d\overline{F}'_i dF'_i\,  e^{-\overline{F}_i F_i} e^{-\overline{F}'_i F'_i} \,J_{FV}(F_f,F'_f,t; F_i,F'_i,0)\, \rho(F_i,F'_i,0)\ .
\end{equation}
$\rho(F_i,F'_i,0)$ is the density matrix at $t=0$ which we take for simplicity to be the vacuum of the unperturbed qubit system
\begin{equation}
	\rho(F_i,F'_i,0)=1\ .
\end{equation}
Hence, we evolve $\rho(F_i,F'_i,0)$ to $\rho(F_f,F'_f,t)$ using the Feynman-Vernon propagator which is given by\footnote{see~\cite{weiss} for the fully bosonic scenario.}
\begin{eqnarray}
	&& J_{FV}(F_f,F'_f,t; F_i,F'_i,0) = \int_{F_i}^{\overline{F}_f} \mathcal{D} \overline{F} \mathcal{D} F \int_{F'_i}^{\overline{F}'_f} \mathcal{D} \overline{F}' \mathcal{D} F' \nonumber \\
	&\times&\exp \left[ \overline{F}_f F(t) + i \int_0^t \left(i \overline{F}(t') \dot{F}(t') - H_{FF}(\overline{F}(t'),F(t'),t')\right) dt' \right. \nonumber \\
	&+&  \left. \overline{F}'(t) F'_f - i \int_0^t \left(-i \dot{\overline{F}}'(t') {F}'(t') - H_{FF}(\overline{F}(t'),F(t'),t')\right) dt'\right] \exp \left[ - S_{FV}(F,F')\right]\ .
\end{eqnarray}
Putting the bosonic degrees of freedom in a heat bath at temperature $1/\beta$, the Feynman-Vernon action $S_{FV}$ takes the form
\begin{eqnarray}
	\exp \left[ - S_{FV}(F,F')\right] &=& \int dB_f dB_i dB'_i \int_{B_i}^{B_f} \mathcal{D} B \int_{B'_i}^{B_f} \mathcal{D} B' \exp \left[i S(B,F) - i S(B',F')\right] \nonumber \\
	&\times& \langle B_i | Z_R^{-1} e^{-\beta \hat{H}_R} | B'_i \rangle
\end{eqnarray}
where 
\begin{equation}\label{eq:SBF}
	S(B,F) = S_{BB}(B,F)+g\,S_{BFF}(B,F)+g\,S_{BBB}(B)+g^2\,S_{BBBB}(B)
\end{equation}
As can be seen from diagram V of Figure~\ref{fig:diagrams}(a), the $S_{BBBB}$ terms do not arise in a context that flips qubits and hence cannot contribute to the boson-qubit coupling at leading order in $g$ in~(\ref{eq:SBF}). They can however arise in the computation of the heat bath terms $Z_R$ and $H_R$ as we shall see shortly. The $S_{BBB}$ terms arise in diagrams III and IV. But once again diagram IV is not involved in the boson-qubit coupling dynamics. This leaves the $S_{BBB}$ terms arising in diagram III only, while the remaining two diagrams, I and II, involve the $S_{BFF}$ terms. Due to the complexity of the computation, we will attempt to simplify things further: we drop diagram III with the following reasoning: considering this diagram in addition to diagram II can only {\em add} interactions between the qubits that may lead to faster but not slower scrambling; and if we can show that diagram II is enough to achieve fast scrambling without considering diagram III, our goal is achieved. On the other hand, if we were to find that the system does {\em not} fast scramble, we would need to come back and add diagram III and check once again. Phrased differently, we are trying to find an {\em upper bound} on the scrambling timescale; and if considering either diagram II {\em or} III demonstrates a timescale bound scaling logarithmically with entropy, we have proved that the system is a fast scrambler. This additional simplification then allows us to drop the $S_{BBB}$ terms as well from~(\ref{eq:SBF}). We then have
\begin{equation}
	S(B,F) \rightarrow S_{BB}(B,F)+g\,S_{BBF}(B,F)
\end{equation}
instead of equation~(\ref{eq:SBF}).
The thermal bath at temperature $1/\beta$ comes in through the expression~\cite{weiss}
\begin{equation}
	\langle B_i | Z_R^{-1} e^{-\beta \hat{H}_{R}} | B'_i \rangle = Z_R^{-1} \int_{B'_i}^{B_i} \mathcal{D} B e^{-S^E_{BB}-g\,S^E_{BBB}-g^2 S^E_{BBBB}}\label{eq:heatbath}
\end{equation}
where $S^E$ is the usual Euclidean action
\begin{equation}
	S^E = \int_0^\beta d\tau\, L^E\ .
\end{equation}
And $Z_R$ is the partition function. For a collection of oscillators of the form
\begin{equation}
	{\dot{x}_k}^2+\lambda_k^2 x_k^2\ ,
\end{equation}
we have
\begin{equation}
	Z_R = \prod_k \frac{1}{2\,\sinh \frac{\beta \lambda_k}{2}}\ .
\end{equation}
We can think of the effect of a small $g$ as shifting the oscillator frequencies $\lambda_k$
\begin{equation}
\lambda_k= \lambda_k^{(0)}+g^2\, \lambda_k^{(2)}+\cdots
\end{equation}
where $\lambda_k^{(0)}$ are the frequencies determined at zeroth order from~(\ref{eq:LBB}). The higher order corrections $\lambda_k^{(i)}$ with $i\geq 2$ involve contributions from diagrams IV and V of Figure~\ref{fig:diagrams}(a) -- that is they come from the $S_{BBB}$ and $S_{BBBB}$ terms of the action. When computing the entanglement entropy in the fermionic sector, we will see that these higher order corrections to $\lambda_k^{(0)}$ drop out since they correspond to disconnected diagrams.

Incorporating all the bosonic and fermionic variables through this Feynman-Vernon process, we finally obtain the effective action
\begin{eqnarray}
	\exp \left[ - S_{FV}(F,F')\right] &=& \frac{1}{2^{2 \left(2 N^2+1\right)} }
	\prod_{j=1}^{N} \left(\mbox{csch}^2\left(\frac{\beta (\lambda_j^1)^2}{2}\right) \sin
	      \left({(\lambda_j^1)^2 t}\right) \mbox{csch}\left({(\lambda_j^1)^2
	      t}\right)\right)^{2 j+1} \nonumber \\
   &&\prod_{j=0}^{N-2} \left(\mbox{csch}^2\left(\frac{1}{2} \beta(\lambda_j^2)^2\right) \sin \left((\lambda_j^2)^2 t\right) \mbox{csch}\left(
   (\lambda_j^2)^2 t\right)\right)^{2 j+1} \nonumber \\
   && \times \exp \left[g^2\sum_{j=1}^{N-2} \mathcal{F}_j^1+g^2\sum_{j=1}^{N} \mathcal{F}_j^2\right]\nonumber \\
   &\equiv & \mathcal{B}(t,N,\beta) \exp \left[g^2\sum_{j=1}^{N-2} \mathcal{F}_j^1+g^2\sum_{j=1}^{N} \mathcal{F}_j^2\right]\label{eq:esfv}
\end{eqnarray}
with
\begin{equation}\label{eq:fmass}
	\lambda^1_j = \frac{j}{3}+O(g^2)\ \ \ ,\ \ \ \lambda^2_j = \frac{j+1}{3}+O(g^2)\ .
\end{equation}
arising from the masses of the bosonic oscillators of equation~(\ref{eq:LBB}); and we define
\begin{eqnarray}
	&& \mathcal{F}_j^i=\int_0^t dt'\int_0^t dt''\frac{1}{{4 \lambda^i_j }}\coth \left(\frac{\beta  \lambda^i_j }{2}\right) \times \nonumber \\
	&& \left(2 \sin \left(\lambda^i_j  t'\right) \sin \left(\lambda^i_j t''\right) \,\omega^{K}_{j}(t')\cdot {\omega'}^{K}_{j}(t'')
	+2 \cos \left(\lambda^i_j  t'\right) \cos \left(\lambda^i_j  t''\right) \,\omega^{K}_{j} (t')\cdot {\omega'}^{K}_{j}(t'')\right. \nonumber \\
	   & & \left.- \sin \left(\lambda^i_j t'\right) \sin \left(\lambda^i_j  t''\right) {\omega'}^{K}_{j} (t')\cdot {\omega'}^{K}_{j}(t'')
	   - \cos \left(\lambda^i_j t'\right) \cos \left(\lambda^i_j  t''\right) {\omega'}^{K}_{j}(t')\cdot {\omega'}^{K}_{j}(t'') \right. \nonumber \\
	   & & \left.- \sin \left(\lambda^i_j  t'\right) \sin \left(\lambda^i_j t''\right) \omega^{K}_{j} (t')\cdot \omega^{K}_{j}(t'')
	   - \cos \left(\lambda^i_j  t'\right) \cos \left(\lambda^i_j  t''\right) \omega^{K}_{j} (t')\cdot \omega^{K}_{j} (t'')\right) \nonumber \\
	   &&+\frac{i}{{2 \lambda^i_j }} \left(\sin \left(\lambda^i_j t'\right) \cos \left(\lambda^i_j  t''\right) \omega^{K}_{j} (t')\cdot {\omega'}^{K}_{j}(t'') 
	   - \cos \left(\lambda^i_j t'\right) \sin \left(\lambda^i_j  t''\right)\omega^{K}_{j} (t')\cdot {\omega'}^{K}_{j}(t'')\right)\ . \label{eq:fij}
\end{eqnarray}
The dot product signifies sum over the $m$ components of the $\omega$s. To leading order in $g^2$, the $\lambda^i_j$s appearing in $\mathcal{F}^i_j$ can be replaced by $j/3$ and $(j+1)/3$ from~(\ref{eq:fmass}). The higher order corrections to the $\lambda^i_j$s do however contribute to the bosonic part $\mathcal{B}(t,N,\beta)$ in~(\ref{eq:esfv}). However, the entire $\mathcal{B}(t,N,\beta)$ factor corresponds to disconnected diagrams and needs to be divided out when writing the density matrix. Hence, the relevant part of the Feynman-Vernon action becomes
\begin{equation}\label{eq:sfvsimple}
	\exp \left[ - S_{FV}(F,F')\right] \rightarrow \exp \left[g^2\sum_{j=1}^{N-2} \mathcal{F}_j^1+g^2\sum_{j=1}^{N} \mathcal{F}_j^2\right]
\end{equation}
with $\lambda^1_j = j/3$ and $\lambda^2_j = (j+1)/3$.
Thus throughout we need not consider contributions from the $L_{BBB}$ and $L_{BBBB}$ terms of the original Lagrangian -- to leading order in $g$.
We then arrive at an effective action for the qubits, equations~(\ref{eq:fij}) and~(\ref{eq:sfvsimple}), that is quartic in the fermions and non-local in time, as expected. 

To proceed further we note that $S_{FV}$ is proportional to $g^2$, and hence we may write
\begin{equation}
 \exp \left[g^2\sum_{j=1}^{N-2} \mathcal{F}_j^1+g^2\sum_{j=1}^{N} \mathcal{F}_j^2\right]\simeq 1 + g^2\sum_{j=1}^{N-2} \mathcal{F}_j^1+g^2\sum_{j=1}^{N} \mathcal{F}_j^2\ ,
\end{equation}
since $g\ll1 $.
This allows us to compute the Feynman-Vernon propagator using the standard technique of a generating functional
\begin{equation}
	J_{FV}(F_f,F'_f,t; F_i,F'_i,0) \simeq \left.\left(1-S_{FV}(\delta/\delta J,\delta/\delta J') \right) G_{FV}(F_f,F'_f,t; F_i,F'_i,0, J, J')\right|_{J,J'=0}
\end{equation}
where we define
\begin{eqnarray}
	&& G_{FV}(F_f,F'_f,t; F_i,F'_i,0) = \int_{F_i}^{\overline{F}_f} \mathcal{D} \overline{F} \mathcal{D} F \int_{F'_i}^{\overline{F}'_f} \mathcal{D} \overline{F}' \mathcal{D} F' \nonumber \\
	&\times&\exp \left[ \overline{F}_f F(t) + i \int_0^t \left(i \overline{F}(t') \dot{F}(t') - H_{FF}(\overline{F}(t'),F(t'),t'\right) dt' \right. \nonumber \\
	&+&  \left. \overline{F}'(t) F'_f - i \int_0^t \left(-i \dot{\overline{F}}'(t') {F}'(t') - H_{FF}(\overline{F}(t'),F(t'),t'\right) dt'\right]  \nonumber \\
	&\times&\exp \left[ i \int_0^t dt' \left(\overline{J}(t') F(t')+\overline{F}(t') J(t')\right) - i \int_0^t dt' \left(\overline{J}'(t') F'(t')+\overline{F}'(t') J(t')\right) \right]
\end{eqnarray}
by introducing the appropriate Grassmanian source terms $J$ and $J'$. We can then proceed by computing $G_{FV}$, and we get (see for example~\cite{weiss})
\begin{eqnarray}
	G_{FV}(F_f,F'_f,t; F_i,F'_i,0) &=&  \left(
	1+e^{-a t} \overline{F}_f\cdot F_i+\overline{F}_f^\cdot H(t)+\overline{H}(t)\cdot F_i
	\right)K(t) \nonumber \\
	&\times& \left(
	1+e^{a t} \overline{F}'_i\cdot F'_f+{H'}^*(t)\cdot\overline{F}'_f + \overline{F}'_i\cdot{\overline{H}'}^*(t)
	\right){K'}^*(t)
\end{eqnarray}
where we have defined
\begin{equation}
	{H}(t) = i \int_0^t dt' J(t') e^{-i a (t-t')}\ \ \ ,\ \ \ \overline{H}(t) = i \int_0^t dt' \overline{J}(t') e^{-i a t'}\ ,
\end{equation}
and
\begin{equation}
	K(t) = \exp \left[-\int_0^t\int_0^t dt' dt'' \overline{J}(t')\cdot J(t'') e^{I a (t'-t'')}\theta(t'-t'')\right]\ .
\end{equation}
In these expressions, $a$ is the mass of the corresponding fermionic mode
\begin{equation}
	a \rightarrow \frac{1}{3} \left(j+\frac{1}{4}\right)\ \ \ \mbox{when $F\rightarrow \eta_{jm}^{I}$ or $F'\rightarrow {\eta'}_{jm}^{I}$}
\end{equation}
and
\begin{equation}
	a \rightarrow \frac{1}{3} \left(j+\frac{3}{4}\right)\ \ \ \mbox{when $F\rightarrow \chi_{jm}^{I}$ or $F'\rightarrow {\chi'}_{jm}^{I}$}	
\end{equation}
from equation~(\ref{eq:LFF}). Putting everything together, we arrive at the density matrix for the qubit system given by
\begin{eqnarray}
	\rho(t) &\simeq & 1+g^2 C^{(1)}(t) + g^2 C^{(2)}_{j_1 m_1 j_2 m_2}(t) \eta'_{f\,j_1 m_1}\cdot\chi'_{f\,j_2 m_2}+ g^2 C^{(3)}_{j_1 m_1 j_2 m_2}(t)\overline{\eta}_{f\,j_1 m_1}\cdot \overline{\chi}_{f\,j_2 m_2} \nonumber \\
	&+& g^2 C^{(4)}_{j_1 m_1 j_2 m_2 j_3 m_3 j_4 m_4}(t) \overline{\eta}_{f\,j_1 m_1}\cdot \overline{\eta}_{f\,j_2 m_2}\cdot\overline{\chi}_{f\,j_3 m_3}\cdot \overline{\chi}_{f\,j_4 m_4} \nonumber \\
	&+& g^2 C^{(5)}_{j_1 m_1 j_2 m_2 j_3 m_3 j_4 m_4}(t)\eta'_{f\,j_1 m_1}\cdot \eta'_{f\,j_2 m_2}\cdot \chi'_{f\,j_3 m_3}\cdot \chi'_{f\,j_4 m_4} \nonumber \\
	&+& g^2 C^{(6)}_{j_1 m_1 j_2 m_2 j_3 m_3 j_4 m_4}(t)\overline{\eta}_{f\,j_1 m_1}\cdot \eta'_{f\,j_2 m_2}\cdot \overline{\chi}_{f\,j_3 m_3}\cdot \chi'_{f\,j_4 m_4}\ ,\label{eq:rho}
\end{eqnarray}
where the $C^{(i)}(t)$ are rather complicated numerical coefficient that we have explicitly computed. The total number of terms in $\rho$ is around $20,000$. However, only $C^{(1)}$ and $C^{(6)}$ will end up contributing to the entanglement entropy. The first involves diagram I of Figure~\ref{fig:diagrams}(a) and includes about $2,000$ combinatorial terms; the last involves diagram II and includes about $1,000$ terms.

\subsection{Entanglement entropy}

Once the density matrix is obtained, computing the Von Neumann entropy is rather straightforward. We first trace over about half of the qubits by integrating out the $\eta$ variables to generate a reduced density matrix
\begin{equation}
	\rho' = \mbox{Tr}_\eta \rho\ .
\end{equation}
This operation takes the integral form over Grassmanian variables
\begin{eqnarray}
	\mbox{Tr}_F \hat{\rho}&& \rightarrow \sum_{n=\pm} \int d \overline{F} dF d {\overline{F}}'  dF' e^{-\overline{F}F} e^{-\overline{F}'F'} \langle n | F \rangle \langle F | \hat{\rho} | F' \rangle  \langle F' | n \rangle \nonumber \\
	&& = \int dF dF' e^{-\overline{F} F} e^{-\overline{F}' F'} \left(\rho + F \rho \overline{F}'\right)
\end{eqnarray}
This gives an expression of the form
\begin{equation}
	\rho' = 1+g^2 K^{(1)}(t) + g^2 K^{(2)}_{j_1 m_1 j_2 m_2}(t)\chi'_{f\,j_1 m_1}\cdot \overline{\chi}_{f\, j_2 m_2}\ .
\end{equation}
$K^{(1)}(t)$ arises from diagram I of Figure~\ref{fig:diagrams}(a) or $C^{(1)}$ in equation~(\ref{eq:rho}); whereas $K^{(2)}_{j_1 m_1 j_2 m_2}(t)$ arises from diagram II or $C^{(6)}$ in equation~(\ref{eq:rho}).
We can then obtain the Von Neumann entropy through 
\begin{equation}
	S(t) = -\mbox{Tr} \hat{\rho}' \ln \hat{\rho}'\ .
\end{equation}
However, there is a subtlety one needs to be careful about. Our computation at one loop renormalizes the qubit wavefunctions. Hence, we need to make sure that the proper counter terms are added so that 
\begin{equation}
	\mbox{Tr}' \rho' = 1\ .
\end{equation}
We write
\begin{equation}
	\rho' = \frac{\overline{\rho}'}{\mathcal{Z}}
\end{equation}
where $\overline{\rho}'$ is the `bare' density matrix computed as a series expansion in $g$
\begin{equation}
	\rho'\rightarrow \overline{\rho}' = \overline{\rho}'_0+g^2 \overline{\rho}'_2+g^4 \overline{\rho}'_4+\cdots
\end{equation}
and $\mathcal{Z}$ is the (wavefunction) renormalization factor
\begin{equation}
	\mathcal{Z} = \mathcal{Z}_0+ g^2 \mathcal{Z}_2+ g^4 \mathcal{Z}_4+\cdots
\end{equation}
and we can see that one needs
\begin{equation}
	\mathcal{Z}_i = \mbox{Tr}' \overline{\rho}'_i\ .
\end{equation}
This then leads to the following expression for the entropy in terms of the bare density matrix
\begin{equation}\label{eq:entropy}
	S \simeq \frac{g^4}{2} \left((\mbox{Tr}' {\overline{\rho}'}_2)^2-\mbox{Tr}' {\overline{\rho}'}^2_2\right)\ .
\end{equation}
Note that the leading contribution of the entanglement entropy arises at order $g^4$ and there is no contribution to this from any part of the density matrix beyond order $g^2$.
Hence, to obtain the entanglement entropy at leading order in $g$, we must compute $\mbox{Tr}' \overline{\rho}'$  and $\mbox{Tr}' {\overline{\rho}'}^2$ -- both of which we have done. However, for the purposes of identifying the scrambling timescale, we have found that looking at the $(\mbox{Tr}' {\overline{\rho}'}_2)^2$ term is enough to demonstrate fast scrambling. Hence, we do not write the very lengthy full result and instead only quote the result for $\mbox{Tr}' \overline{\rho}'$ in Appendix C.

\subsection{Analysis}

In this section, we collect the results of analyzing the evolution of the Von Neumann entropy. We have computed the entire expression given by~(\ref{eq:entropy}) in closed form. The complexity of this computation arises from the combinatorics of the diagrams, {\em i.e.} the many sums over angular momentum modes. The first term of~(\ref{eq:entropy}), $(\mbox{Tr}' {\overline{\rho}'}_2)^2$, involves for example eight such sums and the number of terms scales as $N^8$. The second term, $\mbox{Tr}' {\overline{\rho}'}^2_2$, contains sixteen sums or $\sim N^{16}$ terms. We have tried to use various asymptotic methods to tame these expressions without much success. We are hence unable to present analytical results. But since we have a closed form expression for $S(t)$ -- albeit a long and complicated one -- we can evaluate it numerically. As mentioned earlier, we need to consider large $N$. This makes even the numerical evaluation challenging. For the $\mbox{Tr}' {\overline{\rho}'}^2_2$ term, we have not been able to numerically evaluate the result for $N>4$ in a reasonable amount of time. However, we are able to tackle the $(\mbox{Tr}' {\overline{\rho}'}_2)^2$  term for $N$ up to around $40$ using parallelized algorithms. As we shall see, looking at only the $(\mbox{Tr}' {\overline{\rho}'}_2)^2$ term for such large values of $N$ is enough to demonstrate that the scrambling time indeed scales logarithmically with entropy. Henceforth, we focus then on a numerical analysis of the first term of~(\ref{eq:entropy}), the  $(\mbox{Tr}' {\overline{\rho}'}_2)^2$  term, that arises from diagram I of Figure~\ref{fig:diagrams}(a).

\begin{figure}
	\begin{center}
		 \includegraphics[width=3.0in]{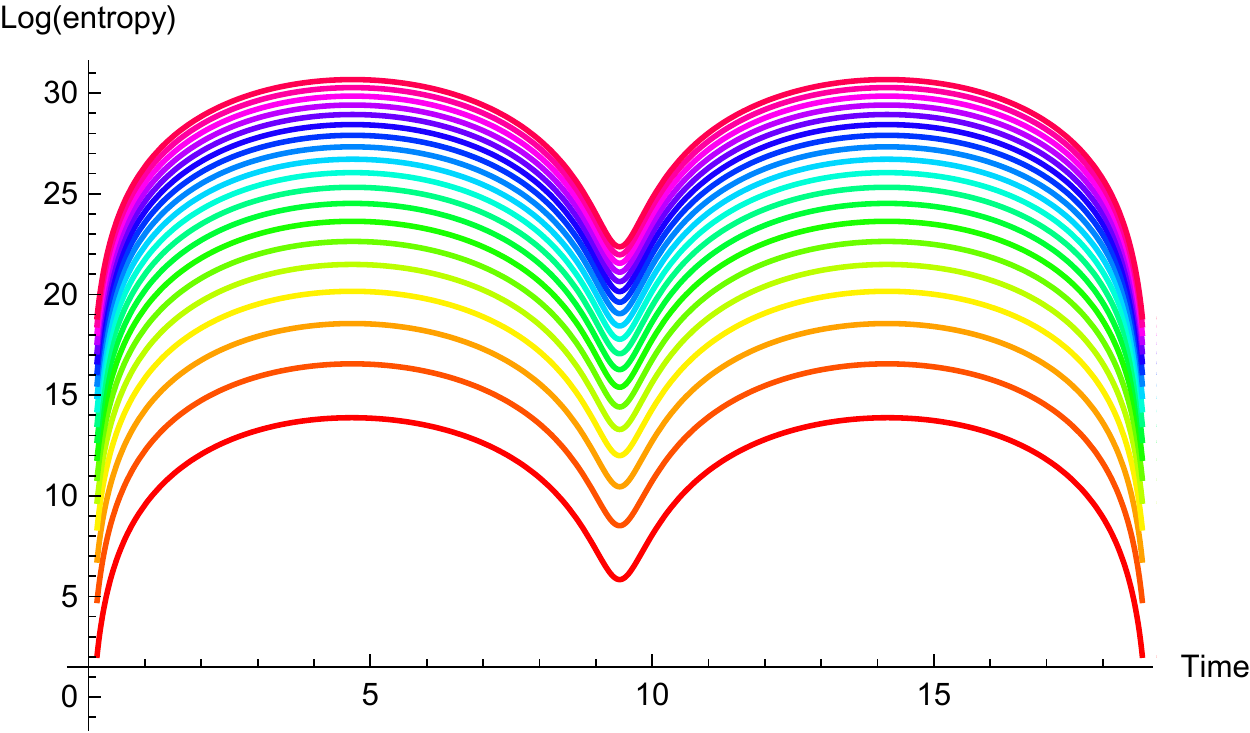}
		 \includegraphics[width=3.0in]{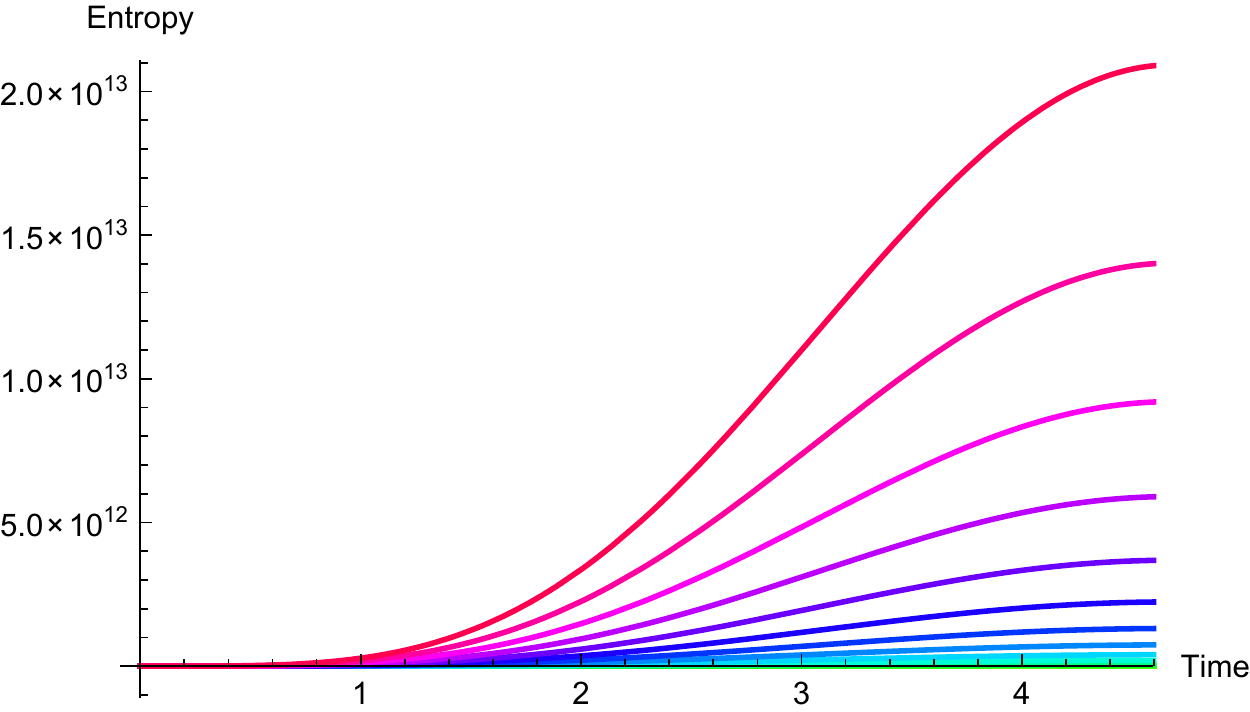}
	\end{center}
	\caption{The Von Neumann entropy as a function of time. $N$ ranges over the odd integers between $N=5$ and $N=41$. We identify an exponential growth period at around $t\sim 2$. Moving upward in each figure, the curves correspond to larger values of $N$.}\label{fig:evolution}
\end{figure}

Figure~\ref{fig:evolution} depicts the time evolution of the Von Neumann entropy due to the first term in~(\ref{eq:entropy}). On the left is a plot of the logarithm of the entropy as a function of time. Noting that the initial state was the vacuum of the unperturbed theory, and that the scheme of obtaining the reduced density matrix involves a rather symmetric partitioning of the full system, we can see that the evolution of the entanglement entropy is not chaotic. We also may be seeing the phenomenon of Poincar\'{e} recurrence since the evolution demonstrates a characteristic cyclic period. The right figure shows a plot of entropy as a function of time zoomed onto a range of timescales for which the entropy evolves exponentially with time. Typically, the process of entropy evolution is characterized by several timescales. The timescales associated with the exponential growth can be identified as the scrambling timescale~\cite{Calabrese:2005in,Liu:2013iza,Brady:2013opa}, which we now focus on. We find it using
\begin{equation}
\tau \sim \frac{S}{\dot{S}}\ .
\end{equation}  

\begin{figure}
	\begin{center}
		\includegraphics[width=4.5in]{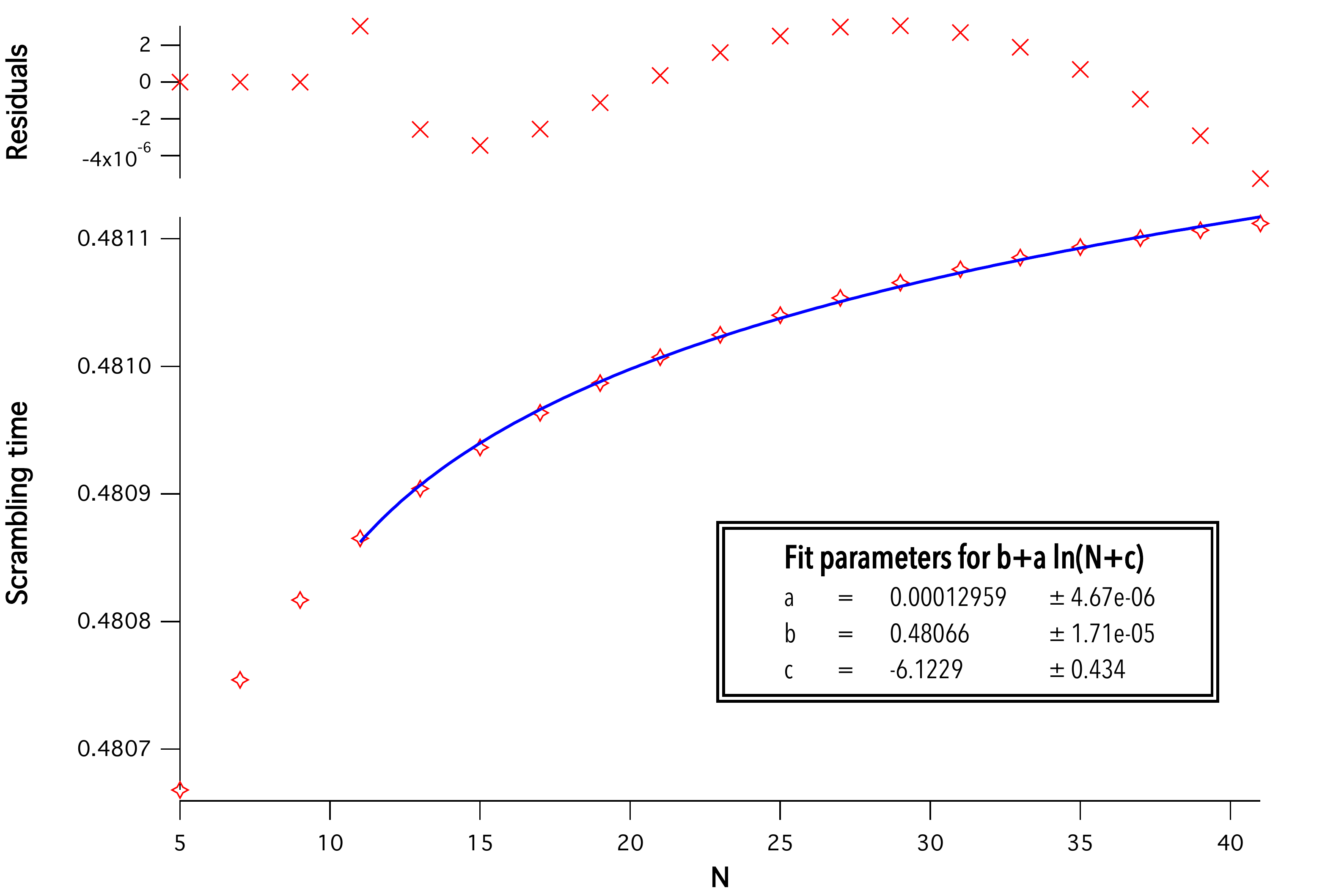}
	\end{center}
	\caption{Scrambling time as a function of $N$. The top graph shows the residuals from the logarithmic fit. Once again, $N$ ranges over the odd integers between $N=5$ and $N=41$.}\label{fig:nplot}
\end{figure}

\begin{figure}
	\begin{center}
		\includegraphics[width=3.0in]{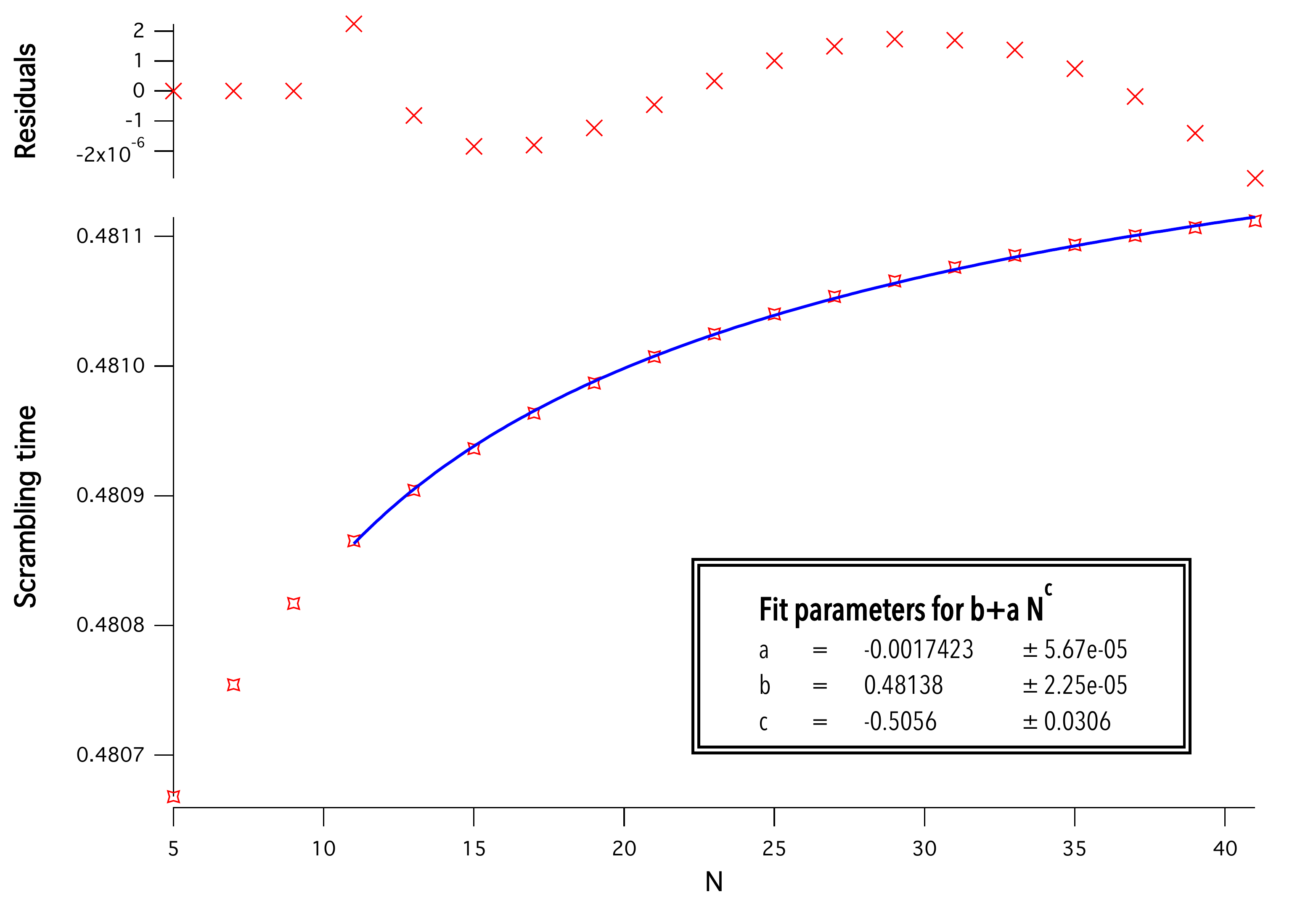}
		\includegraphics[width=3.0in]{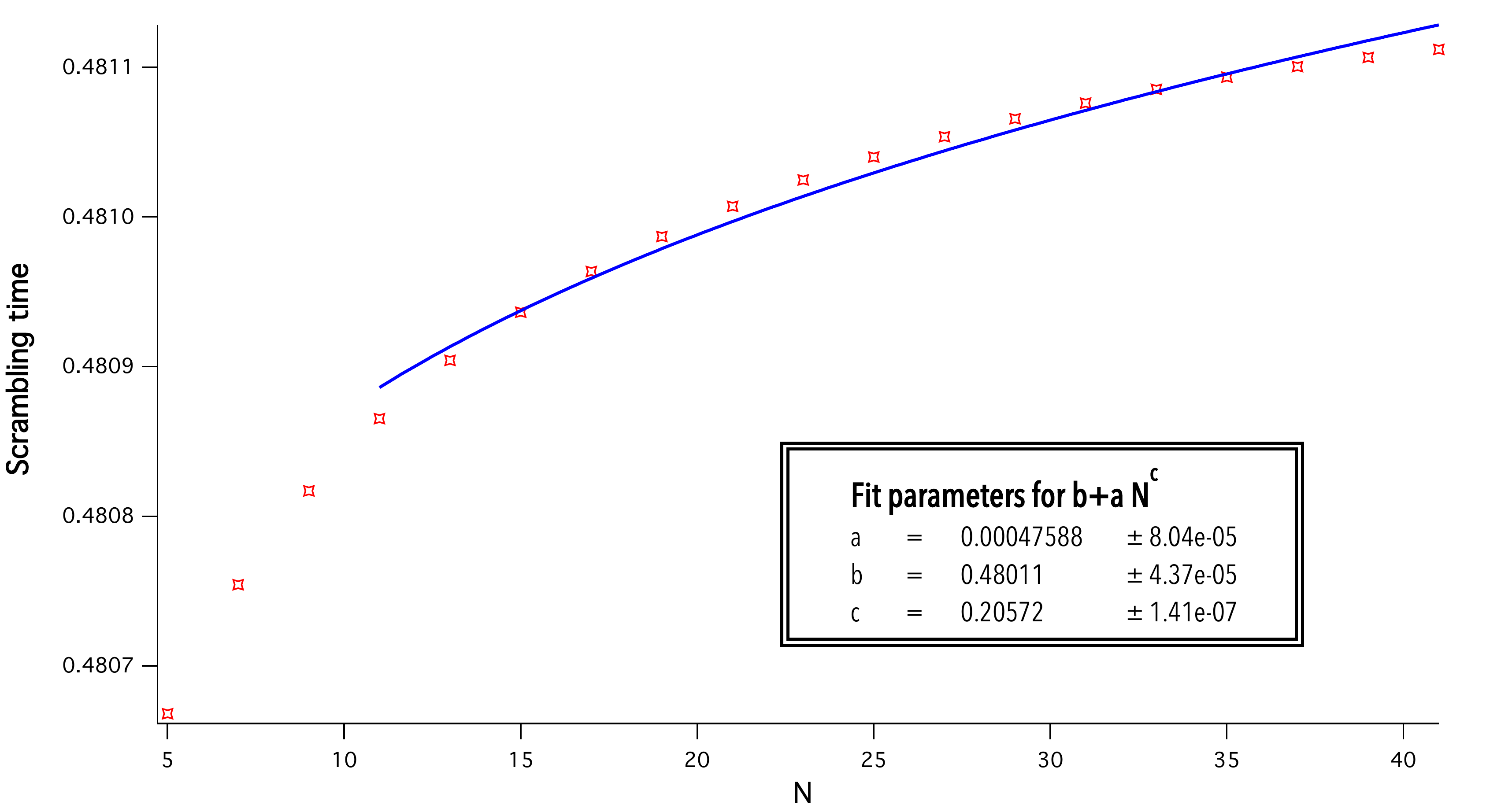}
	\end{center}
	\caption{Power law fits to scrambling time as a function of $N$.}\label{fig:nplot2}
\end{figure}

Figure~\ref{fig:nplot} shows a plot of scrambling time as a function of $N$. Removing the first few small $N$ values, we fit the data to a logarithmic profile and find excellent agreement (with a $\chi^2\sim 1$). 
To contrast this with alternative entropy evolution models, we show two more fits to the data in Figure~\ref{fig:nplot2}. Both are power law fits $\tau \sim N^c$, with the plot on the left having no restrictions on the fit parameters while the plot on the right restricting $c$ to be positive. The plot on the right shows a fit with $\tau\sim N^{0.2}$ -- but, as is apparent, the fit is a very poor one. However, the model on the left gives as good of a fit as the logarithmic model of Figure~\ref{fig:nplot}. It involves a pathological conclusion: $\tau\sim \mbox{constant}-1/\sqrt{N}$. Intuitively, we must have scrambling time scaling at worst as a {\em positive} power of $N$. Note that our analysis truncated away diagram III of Figure~\ref{fig:diagrams}. In this sense, it can only be used to find an {\em upper bound} on scrambling timescale since the additional diagram that we have ignored can only make the scrambling faster. Hence, there is the possibility that our computation is identifying a power law upper bound on scrambling timescale that, over the range of $N$ values we explored, decreases with $N$ but then at some point must and would reverse and increase with $N$ as needed. 
We are then lead to conclude that the data is promising evidence that the BMN matrix model, in the regime of the parameters we have explored, is a fast scrambler as suggested by Figure~\ref{fig:nplot}. Yet we cannot conclusively rule out the possibility that we have instead identified a strange power law upper bound as depicted on the left of Figure~\ref{fig:nplot2}.

\begin{figure}
	\begin{center}
		\includegraphics[width=5.0in]{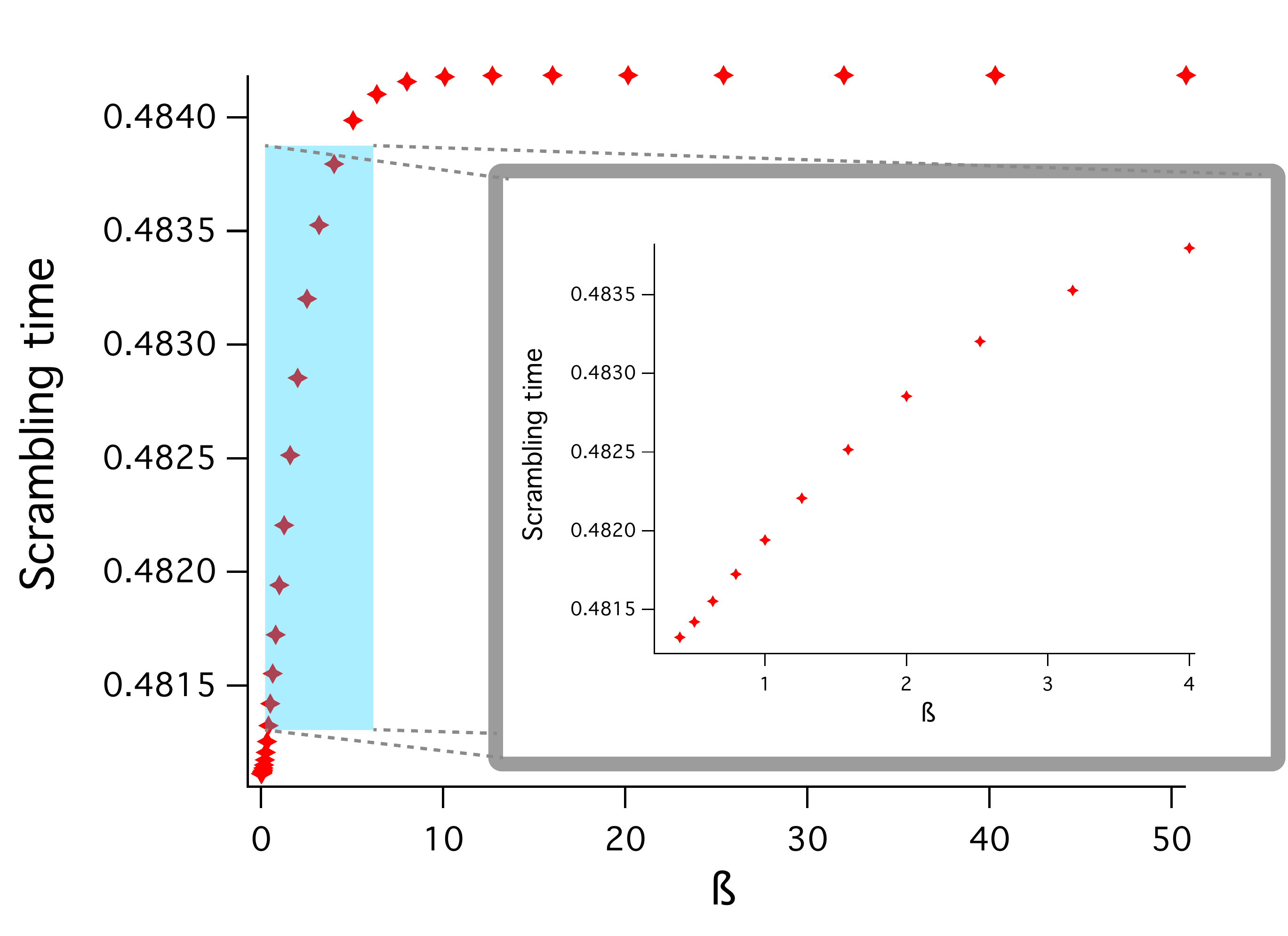}
	\end{center}
	\caption{Scrambling time as a function of $\beta$. High temperature is towards the left of the horizontal axis, low temperature on the right. The inset shows a zoomed out range for low enough temperatures that are still above the IR cutoff scale. For this plot, we have $N=41$.}\label{fig:temp}
\end{figure}
Finally, assuming the promising logarithmic trend of Figure~\ref{fig:nplot}, we look at the temperature dependence of the scrambling time. Figure~\ref{fig:temp} shows the scrambling time as a function of inverse temperature $\beta$. As discussed earlier, the system has a IR cutoff of order one in units of $\mu$. Hence, at low temperatures $\beta\gtrsim 1$, we expect finite size effects to kick in and a freeze out of the heat bath temperature. Correspondingly, we see from the Figure that the scrambling timescale becomes independent of temperature -- that is, there is no dynamics in larger thermal wavelengths than the size of the box in which the system is placed. For low enough temperatures $\beta \lesssim 1$ above the IR cutoff scale, we should however expect $\tau\sim \beta$. The inset in the figure shows a zoomed view of this temperature regime, demonstrating a linear dependence on $\beta$ as expected.

\section{Conclusions and Outlook}\label{sec:conclusion}
\label{sub:conclusion}

Given the connections fast scrambling has to black hole dynamics~\cite{Hayden:2007cs,Sekino:2008he,Susskind:2011ap,Susskind:2013tg}, our result is further circumstantial evidence that BMN Matrix theory is a theory of quantum gravity. However, the mechanism of fast scrambling we identified is also present in the BFSS Lagrangian: a dense network of connections between qubits arising from the commutator of matrices, essentially from the density of the structure constants of equation~(\ref{eq:fs}). Hence, the more general implication of this work is a confirmation of the suggestion by~\cite{Susskind:2011ap} that M-theory-related Matrix theories are indeed fast scramblers. Furthermore, to achieve this fast scrambling, the problem relies on the combinatorial complexities involved in coupling representations of $SU(2)$ -- that is, a dense web of interactions between all spherical harmonic modes -- which in turn makes an asymptotic analytical treatment most challenging. It is interesting to note that the structure constants of equation~(\ref{eq:fs}) which determine the density of the effective qubit-qubit network are peaked for couplings that mix the highest and lowest angular momentum modes -- as if a stringy UV-IR mixing is central to the highly efficient scrambling dynamics. 

The model we employed is very reminiscent of the Matrix black hole of~\cite{Horowitz:1997fr,Banks:1997hz,Banks:1997tn}. In a sense, we have made this original Matrix black hole narrative more concrete: a heat bath associated with the shape of a spherical membrane -- a structure akin to the membrane paradigm of the black hole horizon -- to which one tethers qubits that represent the supergravity degrees of freedom. The equilibrium state would correspond to the two sectors of the model -- the membrane shape degrees of freedom and the qubits -- being in thermal equilibrium. In this work, we artificially arranged the qubits in their ground state and tracked the subsequent evolution towards thermalization. This then naturally allowed us to identify the scrambling timescale. The energy regime we focused on corresponds to non-local fuzzy D2 brane dynamics, strongly coupled yet without gravitational dressing. The implication is that gravitational physics is a red herring in the information scrambling problem: the relevant ingredient is non-local interactions near the black hole horizon\footnote{See also~\cite{Edalati:2012jj} for a similar conclusion from the non-commutative field theory perspective.}. This setup is only one step of exploration in an otherwise very rich system that admits many more interesting black hole-related applications.

For future explorations, there are many interesting venues to pursue. One can complete the full calculation by including all Feynman diagrams to construct the full form of the entanglement entropy as a function of time. While we do not expect that this will add to the punchline of fast scrambling, analyzing the full form of the evolution of the entropy may help identify other interesting timescales that are typically at work in thermalization processes~\cite{Liu:2013iza}. And while we were forced to fall back on a numerical analysis of the final entropy expression, there is the potential for a successful analytical analysis: the asymptotics of the structure constants for $N\rightarrow\infty$ can be worked out using Edmonds' relations and their extensions~\cite{borodin}. The analogue of the thermodynamic limit in our system is indeed the regime $N\rightarrow \infty$; and hence techniques used in traditional statistical systems can perhaps be used to tame the complicated entropy expression.

Black hole evaporation should correspond to qubits freeing themselves from the thermal bath, the membrane at the horizon. The entanglement between Hawking radiation and the black hole then translates to the entanglement between evaporated qubits and membrane-attached qubits. Toy models of qubit-based evaporation mechanisms for black holes have already been explored in the literature~\cite{Avery:2011nb,Giddings:2011ks,Giddings:2012bm}, albeit with too much freedom in choosing arbitrary mechanisms of qubit-qubit interactions. However, this Matrix theory model provides for a concrete string theory-embedded mechanism for determining the details of Matrix black hole evaporation. Along the same line of thought, one can attempt to address the in-fall problem by initially arranging for a small subspace of the matrices away from the spherical configuration and then tracking the merging  of the probe into the membrane structure. Qubits can then once again be used as probes of the evolution of information into the Matrix black hole. Perhaps all this can lead to a better understanding of a horizon firewall~\cite{Almheiri:2012rt,Almheiri:2013hfa,Harlow:2013tf}. BMN Matrix theory also admits other interesting regimes in its parameter space. In particular, spherical five brane vacua can be realized in the theory, along with a similar perturbation problem yet to be explored.

Finally, vacua of BMN Matrix theory have dual holographic spacetimes that are valid geometries in complementary regimes to the one we have focused on~\cite{Lin:2004nb,Ling:2006up}. Our setup corresponds instead to a setting without gravity but with non-local dynamics -- the realm of non-commutative field theories. It would be interesting to develop a more detailed map between qubits or perturbations in the Matrix theory language and corresponding supergravity excitation modes. This would help to directly connect the scrambling dynamics we analyzed to the supergravity black hole scrambling phenomenon. 

\section{Appendices}

\subsection{Appendix A: $W$ coefficients}
	
Here we give a couple of samples for the type of expressions that arise in computing the boson-qubit coupling terms of the action, packaged into what we label as the $W^{(i) K}_{j_1\,m_1\,j_2\,m_2\,j_3}$ coefficients. For example, we have
\begin{eqnarray}
	&&\left(W^{(1) 1}_{j_1\,m_1\,j_2\,m_2\,j_3}\right)_{m_3} = \sqrt{\left(j_1-m_1\right) \left(j_2+m_2\right) \left(j_3-m_3+1\right) \left(j_3-m_3+2\right)} \, \times \nonumber \\ 
&& \left((-1)^{m_2+m_3}
   f^{j_1-\frac{1}{2}\,m_1+\frac{1}{2}\,j_2-\frac{1}{2}\,\frac{1}{2}-m_2}_{j_3+1\,1-m_3}+(-1)^{m_1+m_3}
   f^{j_2-\frac{1}{2}\,m_2-\frac{1}{2}\,j_1-\frac{1}{2}\,-m_1-\frac{1}{2}}_{j_3+1\,1-m_3}\right) \nonumber \\ 
&& +\sqrt{\left(j_1-m_1\right)
   \left(j_2+m_2\right) \left(j_3+m_3+1\right) \left(j_3+m_3+2\right)} \, \times \nonumber \\ 
&& \left((-1)^{m_2}
   f^{j_1-\frac{1}{2}\,m_1+\frac{1}{2}\,j_2-\frac{1}{2}\,\frac{1}{2}-m_2}_{j_3+1\,m_3+1}+(-1)^{m_1}
   f^{j_2-\frac{1}{2}\,m_2-\frac{1}{2}\,j_1-\frac{1}{2}\,-m_1-\frac{1}{2}}_{j_3+1\,m_3+1}\right) \nonumber \\ 
&& +\sqrt{\left(j_1+m_1\right)
   \left(j_2+m_2\right) \left(j_3-m_3+1\right) \left(j_3+m_3+1\right)} \, \times \nonumber \\ 
&& \left((-1)^{m_2+m_3+1}
   f^{j_1-\frac{1}{2}\,m_1-\frac{1}{2}\,j_2-\frac{1}{2}\,\frac{1}{2}-m_2}_{j_3+1\,-m_3}+(-1)^{m_2+1}
   f^{j_1-\frac{1}{2}\,m_1-\frac{1}{2}\,j_2-\frac{1}{2}\,\frac{1}{2}-m_2}_{j_3+1\,m_3} \right. \nonumber \\
&& \left. +(-1)^{m_1+m_3}
   f^{j_2-\frac{1}{2}\,m_2-\frac{1}{2}\,j_1-\frac{1}{2}\,\frac{1}{2}-m_1}_{j_3+1\,-m_3}+(-1)^{m_1}
   f^{j_2-\frac{1}{2}\,m_2-\frac{1}{2}\,j_1-\frac{1}{2}\,\frac{1}{2}-m_1}_{j_3+1\,m_3}\right) \nonumber \\ 
&& +\sqrt{\left(j_1-m_1\right)
   \left(j_2-m_2\right) \left(j_3-m_3+1\right) \left(j_3+m_3+1\right)} \, \times \nonumber \\ 
&& \left((-1)^{m_2+m_3+1}
   f^{j_1-\frac{1}{2}\,m_1+\frac{1}{2}\,j_2-\frac{1}{2}\,-m_2-\frac{1}{2}}_{j_3+1\,-m_3}+(-1)^{m_2+1}
   f^{j_1-\frac{1}{2}\,m_1+\frac{1}{2}\,j_2-\frac{1}{2}\,-m_2-\frac{1}{2}}_{j_3+1\,m_3} \right. \nonumber \\
&& \left. +(-1)^{m_1+m_3}
   f^{j_2-\frac{1}{2}\,m_2+\frac{1}{2}\,j_1-\frac{1}{2}\,-m_1-\frac{1}{2}}_{j_3+1\,-m_3}+(-1)^{m_1}
   f^{j_2-\frac{1}{2}\,m_2+\frac{1}{2}\,j_1-\frac{1}{2}\,-m_1-\frac{1}{2}}_{j_3+1\,m_3}\right) \nonumber \\ 
&& +\sqrt{\left(j_1+m_1\right)
   \left(j_2-m_2\right) \left(j_3+m_3+1\right) \left(j_3+m_3+2\right)} \, \times \nonumber \\ 
&& \left((-1)^{m_2+m_3}
   f^{j_1-\frac{1}{2}\,m_1-\frac{1}{2}\,j_2-\frac{1}{2}\,-m_2-\frac{1}{2}}_{j_3+1\,-m_3-1}+(-1)^{m_1+m_3}
   f^{j_2-\frac{1}{2}\,m_2+\frac{1}{2}\,j_1-\frac{1}{2}\,\frac{1}{2}-m_1}_{j_3+1\,-m_3-1}\right) \nonumber \\ 
&& +\sqrt{\left(j_1+m_1\right)
   \left(j_2-m_2\right) \left(j_3-m_3+1\right) \left(j_3-m_3+2\right)} \, \times \nonumber \\ 
&& \left((-1)^{m_2}
   f^{j_1-\frac{1}{2}\,m_1-\frac{1}{2}\,j_2-\frac{1}{2}\,-m_2-\frac{1}{2}}_{j_3+1\,m_3-1}+(-1)^{m_1}
   f^{j_2-\frac{1}{2}\,m_2+\frac{1}{2}\,j_1-\frac{1}{2}\,\frac{1}{2}-m_1}_{j_3+1\,m_3-1}\right)
\end{eqnarray}
or
\begin{eqnarray}
&& W^{(3) 3}_{j_1\,m_1\,j_2\,m_2\,j_3} = \sqrt{\left(j_3+1\right) \left(j_3+2\right) \left(j_1+m_1+1\right) \left(j_2+m_2\right)} \, \times \nonumber \\ 
&& \left((-1)^{m_2}
   f^{j_1+\frac{1}{2}\,m_1+\frac{1}{2}\,j_2-\frac{1}{2}\,\frac{1}{2}-m_2}_{j_3+1\,1}+(-1)^{m_1}
   f^{j_2-\frac{1}{2}\,m_2-\frac{1}{2}\,j_1+\frac{1}{2}\,-m_1-\frac{1}{2}}_{j_3+1\,1}\right) \nonumber \\ 
&& +\sqrt{\left(j_1-m_1+1\right)
   \left(j_2+m_2\right)} \, \times \nonumber \\ 
&& \left((-1)^{m_2} f^{j_1+\frac{1}{2}\,m_1-\frac{1}{2}\,j_2-\frac{1}{2}\,\frac{1}{2}-m_2}_{j_3+1\,0}+j_3
   (-1)^{m_2} f^{j_1+\frac{1}{2}\,m_1-\frac{1}{2}\,j_2-\frac{1}{2}\,\frac{1}{2}-m_2}_{j_3+1\,0} \right. \nonumber \\
&& \left. +(-1)^{m_1+1}
   f^{j_2-\frac{1}{2}\,m_2-\frac{1}{2}\,j_1+\frac{1}{2}\,\frac{1}{2}-m_1}_{j_3+1\,0}+j_3 (-1)^{m_1+1}
   f^{j_2-\frac{1}{2}\,m_2-\frac{1}{2}\,j_1+\frac{1}{2}\,\frac{1}{2}-m_1}_{j_3+1\,0}\right) \nonumber \\ 
&& +\sqrt{\left(j_1+m_1+1\right)
   \left(j_2-m_2\right)} \, \times \nonumber \\ 
&& \left((-1)^{m_2+1} f^{j_1+\frac{1}{2}\,m_1+\frac{1}{2}\,j_2-\frac{1}{2}\,-m_2-\frac{1}{2}}_{j_3+1\,0}+j_3
   (-1)^{m_2+1} f^{j_1+\frac{1}{2}\,m_1+\frac{1}{2}\,j_2-\frac{1}{2}\,-m_2-\frac{1}{2}}_{j_3+1\,0} \right. \nonumber \\
&& \left. +(-1)^{m_1}
   f^{j_2-\frac{1}{2}\,m_2+\frac{1}{2}\,j_1+\frac{1}{2}\,-m_1-\frac{1}{2}}_{j_3+1\,0}+j_3 (-1)^{m_1}
   f^{j_2-\frac{1}{2}\,m_2+\frac{1}{2}\,j_1+\frac{1}{2}\,-m_1-\frac{1}{2}}_{j_3+1\,0}\right) \nonumber \\ 
&& +\sqrt{\left(j_3+1\right)
   \left(j_3+2\right) \left(j_1-m_1+1\right) \left(j_2-m_2\right)} \, \times \nonumber \\ 
&& \left((-1)^{m_2+1}
   f^{j_1+\frac{1}{2}\,m_1-\frac{1}{2}\,j_2-\frac{1}{2}\,-m_2-\frac{1}{2}}_{j_3+1\,-1}+(-1)^{m_1+1}
   f^{j_2-\frac{1}{2}\,m_2+\frac{1}{2}\,j_1+\frac{1}{2}\,\frac{1}{2}-m_1}_{j_3+1\,-1}\right)
\end{eqnarray}

In writing $W^{(i) K}_{j_1\,m_1\,j_2\,m_2\,j_3}$, we have $i=1,\ldots, 4$ and $K=1,\cdots, 6$. For $K=3$ and $K=6$, the $W^{(i) K}_{j_1\,m_1\,j_2\,m_2\,j_3}$ is a scalar, but for other $K$s it is a vector in the $m_3$ index. As shown in Appendix B, the $f$s are $N$ independent. The interactions packaged in the $W$s are essentially a packaging of the combinatorics of combining $SU(2)$ representations. 

\subsection{Appendix B: Structure constants}

\begin{figure}
	\begin{center}
		\includegraphics[width=4.0in]{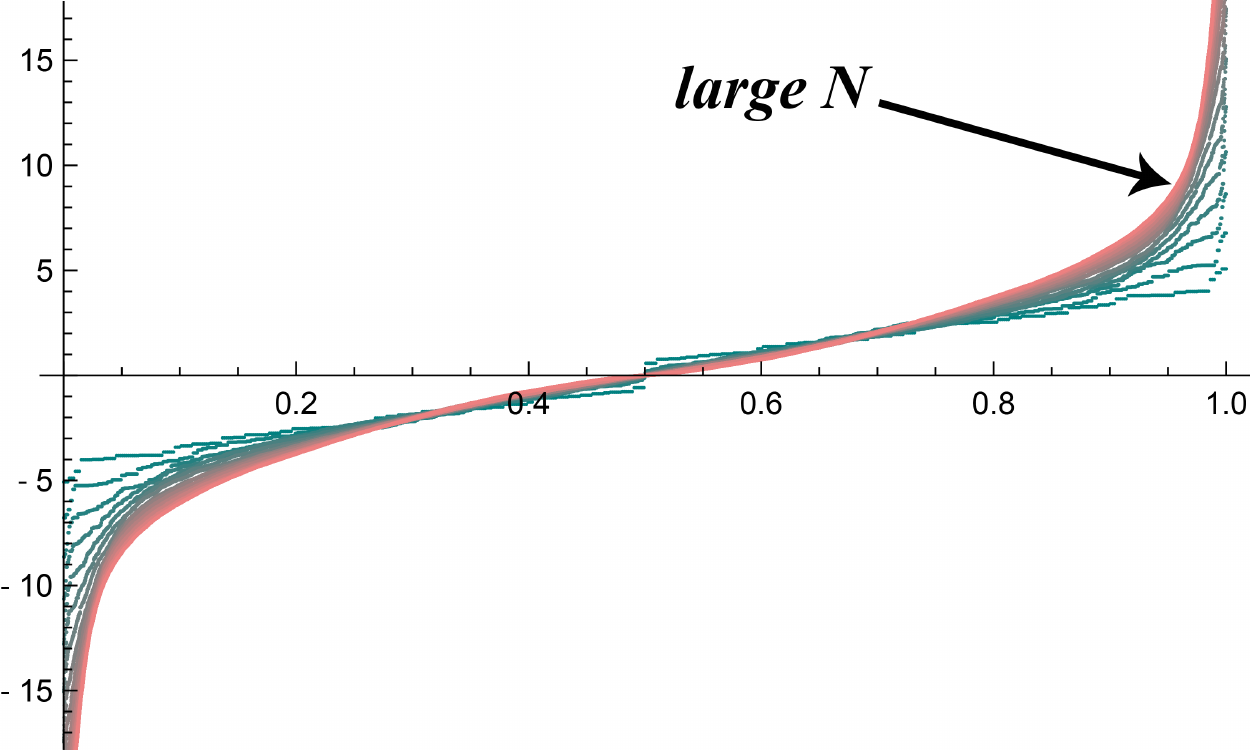}
	\end{center}
	\caption{Plot of $f^{j_1\,m_1\,j_2\,m_2}_{j_3\,m_3}$ versus all allowed values of $(j_1,m_1,j_2,m_2,j_3,m_3)$. The more jagged curves correspond to smaller values of $N$.}\label{fig:fplots}
\end{figure}

In this appendix, we show that the structure constants $f^{j_1\,m_1\,j_2\,m_2}_{j_3\,m_3}$ are $N$ independent. Figure~\ref{fig:fplots} shows a plot of $f^{j_1\,m_1\,j_2\,m_2}_{j_3\,m_3}$ for all values of its indices. The horizontal axis is scaled from $0$ to $1$ to map onto the various possibilities of $(j_1,m_1,j_2,m_2,j_3,m_3)$ -- sorted according to the value of the corresponding $f^{j_1\,m_1\,j_2\,m_2}_{j_3\,m_3}$. The different curves correspond to different values of $N$, from $N=5$ to $N=21$. We see a clear convergence pattern, with the maximum and minimum values of $f$ capped off about $\pm 18$, independent of $N$. Indeed, the asymptotic form of this profile for large $N$ can be obtained in closed form through Edmonds formula~\cite{borodin}. The peaks of $f$ arise when two of the $j$s are maximized and the third is minimized. For example, for $N=11$, maxima arise around $(j_1,m_1,j_2,m_2,j_3,m_3)=(1,0,10,-10,10,10)$ and permutations of these $(j_1,m_1)$, $(j_2,m_2)$, and $(j_3,m_3)$. This implies that the predominant contribution in our computation comes from dynamics that mixes the highest and lowest angular momentum modes.
The $N$ independence of the $f$s inspires the definition of the coupling constant given in~(\ref{eq:g}) for large $N$.

\subsection{Appendix C: Entropy evolution}

In this appendix, we write the full expression for $\mbox{Tr}' \overline{\rho}'$, the square of which gives us part of the entropy as a function of time. 
\begin{eqnarray}
	\mbox{Tr}' \overline{\rho}' &=& \sum_{i=1}^2 \sum_{j=1}^{N+2 (i-2)} \left\{\frac{}{}\right. \nonumber \\
	 && \left[ \sum_{j_1=1/2}^{N-1/2} \sum_{m_1} W^{(1)K}_{j_1m_1\,j_1m_1;j}\cdot W^{(1)K}_{j_1m_1\,j_1m_1;j}  + \sum_{j_2=1/2}^{N-3/2} \sum_{m_2} W^{(2)K}_{j_2m_2\,j_2m_2;j}\cdot W^{(2)K}_{j_2m_2\,j_2m_2;j}\right] \times \nonumber \\
	 && \frac{24 \coth \left(\frac{\beta  \lambda ^i_j}{2}\right) \sin
	   ^2\left(\frac{t \lambda^i_j}{2}\right)}{\left(\lambda
	   ^i\right)_j^3} \nonumber \\
	   && + \sum_{j_1=1/2}^{N-1/2} \sum_{m_1} \sum_{j_2=1/2}^{N-1/2} \sum_{m_2}
	   \frac{W^{(3)K}_{j_2m_2\,j_1m_1;j}\cdot
	      W^{(4)K}_{j_2m_2\,j_1m_1;j}}{\lambda ^i_j \left(\left(\lambda
	      ^i\right)_j^2-\left(a_{j_2}+b_{j_1}\right){}^2\right){}^2} \times \nonumber \\
	   && 4 \coth \left(\frac{\beta  \lambda ^i_j}{2}\right) \left[
	   4 \left(a_{j_2}+b_{j_1}\right) \lambda ^i_j \sin \left(t
	      \left(a_{j_2}+b_{j_1}\right)\right) \sin \left(t \lambda
	      ^i_j\right) \right. \nonumber \\
		  && \left. -2 \left(\left(a_{j_2}+b_{j_1}\right){}^2+\left(\lambda
	      ^i\right)_j^2\right) \left(1-\cos \left(t
	      \left(a_{j_2}+b_{j_1}\right)\right) \cos \left(t \lambda
	      ^i_j\right)\right)
	   \right] \nonumber \\
	  && +\sum_{j_1=1/2}^{N-1/2} \sum_{m_1} \sum_{j_2=1/2}^{N-1/2} \sum_{m_2} \sum_{j_3=1/2}^{N-1/2} \sum_{m_3}
	  \frac{2 W^{(3)K}_{j_3m_3\,j_1m_1;j}\cdot
	     W^{(4)K}_{j_2m_2\,j_1m_1;j}}{\lambda ^i_j \left(\left(\lambda
	     ^i\right)_j^2-\left(a_{j_2}+b_{j_1}\right){}^2\right){}^2} \times \nonumber \\
	  && \left[
	  a_{j_2} \left(b_{j_1} \coth \left(\frac{\beta  \lambda
	     ^i_j}{2}\right)-\lambda ^i_j\right)+a_{j_3}
	     \left(\left(a_{j_2}+b_{j_1}\right) \coth \left(\frac{\beta  \lambda
	     ^i_j}{2}\right)-\lambda ^i_j\right) \right. \nonumber \\
		 && \left. +b_{j_1}^2 \coth
	     \left(\frac{\beta  \lambda ^i_j}{2}\right)-2 b_{j_1} \lambda
	     ^i_j+\left(\lambda ^i\right)_j^2 \coth \left(\frac{\beta  \lambda
	     ^i_j}{2}\right)
	  \right]	  \nonumber \\
	&& + \sum_{j_1=1/2}^{N-1/2} \sum_{m_1} \sum_{j_2=1/2}^{N-1/2} \sum_{m_2} \sum_{j_3=1/2}^{N-1/2} \sum_{m_3} 
	\frac{W^{(3)K}_{j_3m_3\,j_1m_1;j}\cdot
	   W^{(4)K}_{j_2m_2\,j_1m_1;j}}{2 \lambda ^i_j
	   \left(\left(\lambda
	   ^i\right)_j^2-\left(a_{j_2}+b_{j_1}\right){}^2\right)
	   \left(\left(\lambda
	   ^i\right)_j^2-\left(a_{j_3}+b_{j_1}\right){}^2\right)}  \times \nonumber \\
	&& \left[
	4 e^{-i t \left(a_{j_3}-a_{j_2}\right)} \left(a_{j_2} \left(b_{j_1}
	   \coth \left(\frac{\beta  \lambda ^i_j}{2}\right)-\lambda
	   ^i_j\right)+a_{j_3} \left(\left(a_{j_2}+b_{j_1}\right) \coth
	   \left(\frac{\beta  \lambda ^i_j}{2}\right)-\lambda
	   ^i_j\right) \right. \right. \nonumber \\
	   && \left. \left. +b_{j_1}^2 \coth \left(\frac{\beta  \lambda
	   ^i_j}{2}\right)-2 b_{j_1} \lambda ^i_j+\left(\lambda
	   ^i\right)_j^2 \coth \left(\frac{\beta  \lambda
	   ^i_j}{2}\right)\right) \right. \nonumber \\
	   && \left. -4 e^{i t \left(a_{j_2}+b_{j_1}\right)}
	   \mbox{csch}\left(\frac{\beta  \lambda ^i_j}{2}\right)
	   \left(a_{j_2} \left(b_{j_1} \cos \left(t \lambda ^i_j-\frac{1}{2}
	   i \beta  \lambda ^i_j\right)-i \lambda ^i_j \sin \left(t \lambda
	   ^i_j-\frac{1}{2} i \beta  \lambda ^i_j\right)\right) \right. \right. \nonumber \\
	   && \left. \left. +a_{j_3}
	   \left(\left(a_{j_2}+b_{j_1}\right) \cos \left(t \lambda
	   ^i_j-\frac{1}{2} i \beta  \lambda ^i_j\right)-i \lambda ^i_j
	   \sin \left(t \lambda ^i_j-\frac{1}{2} i \beta  \lambda
	   ^i_j\right)\right) \right. \right. \nonumber \\
	   && \left. \left. -2 i b_{j_1} \lambda ^i_j \sin \left(t \lambda
	   ^i_j-\frac{1}{2} i \beta  \lambda ^i_j\right)+b_{j_1}^2 \cos
	   \left(t \lambda ^i_j-\frac{1}{2} i \beta  \lambda
	   ^i_j\right)+\left(\lambda ^i\right)_j^2 \cos \left(t \lambda
	   ^i_j-\frac{1}{2} i \beta  \lambda ^i_j\right)\right) \right. \nonumber \\
	   && \left. -4 e^{-i t
	   \left(a_{j_3}+b_{j_1}\right)} \mbox{csch}\left(\frac{\beta  \lambda
	   ^i_j}{2}\right) \left(a_{j_2} \left(b_{j_1} \cos \left(t \lambda
	   ^i_j+\frac{1}{2} i \beta  \lambda ^i_j\right)+i \lambda ^i_j
	   \sin \left(t \lambda ^i_j+\frac{1}{2} i \beta  \lambda
	   ^i_j\right)\right) \right. \right. \nonumber \\
	   && \left. \left. +a_{j_3} \left(\left(a_{j_2}+b_{j_1}\right) \cos
	   \left(t \lambda ^i_j+\frac{1}{2} i \beta  \lambda ^i_j\right)+i
	   \lambda ^i_j \sin \left(t \lambda ^i_j+\frac{1}{2} i \beta
	   \lambda ^i_j\right)\right) \right. \right. \nonumber \\
	   && \left. \left. +2 i b_{j_1} \lambda ^i_j \sin \left(t
	   \lambda ^i_j+\frac{1}{2} i \beta  \lambda ^i_j\right)+b_{j_1}^2
	   \cos \left(t \lambda ^i_j+\frac{1}{2} i \beta  \lambda
	   ^i_j\right) \right. \right. \nonumber \\
	   && \left. \left. +\left(\lambda ^i\right)_j^2 \cos \left(t \lambda
	   ^i_j+\frac{1}{2} i \beta  \lambda ^i_j\right)\right)
	\right]\left.\frac{}{}\right\}
\end{eqnarray}
The terms involving four or six sums arise from diagram I of Figure~\ref{fig:diagrams}(a) while the rest arise from diagram II. We have also defined the masses
\begin{equation}
	\lambda^1_j = \frac{j}{3}\ \ \ ,\ \ \ \lambda^2_j = \frac{j+1}{3}
\end{equation}
\begin{equation}
	a_j \rightarrow \frac{1}{3} \left(j+\frac{1}{4}\right)\ \ \ ,\ \ \ b_j \rightarrow \frac{1}{3} \left(j+\frac{3}{4}\right)\ .
\end{equation}

\section{Acknowledgments}

This work was supported by NSF grant number PHY-0968726, and a gift of two Tesla GPU cards from nVidia.


\providecommand{\href}[2]{#2}\begingroup\raggedright\endgroup

\end{document}